\documentclass[12pt]{article}
\usepackage{epsf}

\usepackage{epsfig,graphics}
\usepackage {graphicx}
\usepackage {epsfig}
\usepackage{subcaption}
\usepackage {tabularx} 
\usepackage{rotate}	
\usepackage{slashed}
\usepackage{bbm}
\usepackage{color}
\usepackage{tikz}
\usetikzlibrary{decorations.pathmorphing} 
\usepackage{amsmath}
\usepackage{amsfonts}
\usepackage{amssymb}
\usepackage{graphicx}
\usepackage{cite}

\usepackage{fancyhdr}
\usepackage{hyperref}
\usepackage{diagbox}

\newcommand{\bmat}{\left(\begin{array}}
\newcommand{\emat}{\end{array}\right)}

\def\yzero{\smash{\hbox{$y\kern-4pt\raise1pt\hbox{${}^\circ$}$}}}

\def\beq{\begin{equation}}
\def\eeq{\end{equation}}
\def\beqa{\begin{eqnarray}}
\def\eeqa{\end{eqnarray}}

\def\-{\hphantom{-}}

\def\s2{\frac{1}{\sqrt2}}

\def\Dsl{\,\raise.15ex\hbox{/}\mkern-13.5mu D} 
\def\IZ{Z\kern-.4em  Z}

\def\be{\begin{equation}}
\def\ee{\end{equation}}
\def\bea{\begin{eqnarray}}
\def\eea{\end{eqnarray}}
\def\bes{\begin{subequations}}
\def\ees{\end{subequations}}

\newcommand{\mathtitles}[1]{\texorpdfstring{$#1$}{Lg}}
\catcode`\@=11   
\newdimen\@rotdimen
\newbox\@rotbox  

\def\@vspec#1{\special{ps:#1}}
\def\@rotstart#1{\@vspec{gsave currentpoint currentpoint translate
   #1 neg exch neg exch translate}}
\def\@rotfinish{\@vspec{currentpoint grestore moveto}}
\def\@rotr#1{\@rotdimen=\ht#1\advance\@rotdimen by\dp#1%
   \hbox to\@rotdimen{\hskip\ht#1\vbox to\wd#1{\@rotstart{90 rotate}%
   \box#1\vss}\hss}\@rotfinish}
\def\@rotl#1{\@rotdimen=\ht#1\advance\@rotdimen by\dp#1%
   \hbox to\@rotdimen{\vbox to\wd#1{\vskip\wd#1\@rotstart{270 rotate}%
   \box#1\vss}\hss}\@rotfinish}%
\def\@rotu#1{\@rotdimen=\ht#1\advance\@rotdimen by\dp#1%
   \hbox to\wd#1{\hskip\wd#1\vbox to\@rotdimen{\vskip\@rotdimen
   \@rotstart{-1 dup scale}\box#1\vss}\hss}\@rotfinish}%
\def\@rotf#1{\hbox to\wd#1{\hskip\wd#1\@rotstart{-1 1 scale}%
   \box#1\hss}\@rotfinish}%
\def\rotate{\@ifnextchar[{\@rotate}{\@rotate[l]}}
\def\@rotate[#1]#2{\setbox\@rotbox=\hbox{#2}\@nameuse{@rot#1}\@rotbox}

\catcode`\@=12

\topmargin
-1.5cm
\textwidth
15.5cm
\textheight
23.5cm
\oddsidemargin
0.7cm
\evensidemargin
0.7cm

\begin{document}

\makeatletter
\@addtoreset{equation}{section}
\makeatother
\renewcommand{\theequation}{\thesection.\arabic{equation}}
\pagestyle{empty}
\vspace{-0.2cm}
\rightline{ IFT-UAM/CSIC-25-69}
\vspace{0.5cm}
\begin{center}
\LARGE{ Moduli Self-Fixing }
\\[8mm]
\large{Gonzalo F.~Casas,$^\diamondsuit$  Luis E.~Ib\'a\~nez$^{\clubsuit  \diamondsuit}$}
\\[6mm]
\small{$^\clubsuit$  Departamento de F\'{\i}sica Te\'orica \\ Universidad Aut\'onoma de Madrid,
Cantoblanco, 28049 Madrid, Spain}  \\[5pt]
$^\diamondsuit$  {Instituto de F\'{\i}sica Te\'orica UAM-CSIC, c/ Nicolas Cabrera 13-15, 28049 Madrid, Spain} 
\\[6mm]
\small{\bf Abstract} \\ [6mm]
\end{center}
\begin{center}
\begin{minipage}[h]{15.22cm}

In Quantum Gravity (QG), large moduli values lead to towers of exponentially light states, making the QG cut-off field-dependent. In 4D supersymmetric (SUSY) theories, this cut-off is set by the species scale $\Lambda(z_i, \bar{z}_i)$, where $z_i$ are complex moduli.
We argue that in GKP-like 4D no-scale vacua, accounting for this field dependence generates 
one-loop, positive-definite potentials for the otherwise unfixed moduli, with local Minkowski minima at the \textit{desert points} in moduli space with $z_i \sim \mathcal{O}(1)$. As these no-scale moduli grow, a dS plateau emerges and, for larger moduli, 
the potential runs away to zero, consistent with Swampland expectations.
 This may have important consequences for the moduli fixing problem. In particular non-perturbative superpotentials may not 
 be necessary for fixing the K\"ahler moduli in a Type IIB setting. 
 Although one loses control over non-perturbative corrections, we argue that modular
 invariance (more generally, dualities)  of the species scale may give us information about the behaviour at small moduli in some simple cases.
 We illustrate this mechanism in a Type II 4D $\mathbb{Z}_2 \times \mathbb{Z}_2$ toroidal orientifold, where in some cases, modular invariance can give us control over the non-perturbative corrections. The vanishing cosmological constant reflects a delicate cancellation between IR and UV contributions at the minimum.  The models may be completed by the addition of intersecting D6-branes, yielding a 3-generation MSSM (RR tadpole free) model with some (but not all) closed string moduli fixed in Minkowski, with the complex structure moduli 
 fixed at the \textit{desert points}.

\end{minipage}
\end{center}
\newpage
\setcounter{page}{1}
\pagestyle{plain}
\renewcommand{\thefootnote}{\arabic{footnote}}
\setcounter{footnote}{0}



\tableofcontents

 %
 	
\section{Introduction}
\label{introduction}

One of the hardest challenges in making contact between string theory and the observed world is the problem of moduli fixing. 
Important progress was achieved by realizing that generic string compactifications (e.g., Type II CY orientifolds) 
have antisymmetric fields with fluxes that can generate a potential for fixing many moduli \cite{Douglas:2006es,Becker:2006dvp,Marchesano:2007de,Denef:2008wq,Quevedo:2014xia,McAllister:2023vgy,Gukov:1999ya}. This appeared in particular in the 
GKP kind of Type IIB vacua in which  RR and NS fluxes may fix, in principle, all complex structure (c.s.), and complex dilaton moduli
in a non-SUSY Minkowski vacuum (at the tree level) \cite{Giddings:2001yu}.  Still in this setting, all the 'no-scale moduli', in this case the K\"ahler moduli,
remain unstabilized. Similar Type IIA models may be constructed in which it is the c.s. moduli which are the no-scale moduli and
remain undetermined \cite{Derendinger:2004jn,Villadoro:2005cu,DeWolfe:2005uu,Camara:2005dc} \footnote{We ignore the gravitational back-reaction, which induces a warping effect in Type IIB and is slightly more difficult to include in Type IIA, since they do not play a relevant role in our discussion.}.  In order to fix the remaining K\"ahler moduli in e.g., IIB, it was proposed in KKLT 
to consider instanton (or gaugino condensation) generated superpotentials $W(T_i)$ \cite{Kachru:2003aw}. This indeed can fix the remaining moduli but at the cost of
landing in an AdS vacuum (SUSY in the KKLT case, non-SUSY in the LV scenario \cite{Balasubramanian:2005zx,Conlon:2005ki}).  By the addition of anti-D3 branes at the tip of a KS throat \cite{Kachru:2003aw},
one can finally get back to Minkowski or even dS by tuning the superpotential $W_0$ and the fluxes at the throat. Other mechanisms involve the introduction of D7 fluxes \cite{Burgess:2003ic,Saltman:2004sn}.
A lot of work has been done developing this general program (see \cite{Grana:2005jc,Blumenhagen:2006ci,Denef:2007pq,Ibanez:2012zz,McAllister:2023vgy} for a review), which may be considered at least partially successful but, admittedly, 
slightly cumbersome. One of the most complicated aspects is to keep all approximations under control so that parametrically 
we remain in a perturbative regime both in an $\alpha '$ and loop expansions, i.e., we want the minima to sit at sufficiently large 
moduli vevs.

In the present paper, starting from the first step, i.e., a GKP-like \cite{Giddings:2001yu} no-scale setting in which some no-scale moduli remain massless,
we compute one-loop corrections to the potential of these initially massless fields. We find that the potential obtained at large moduli
goes like $V\sim g^2 m_{3/2}^2M_{\rm P}^2$, where $g$ denotes the effective coupling appearing in the one-loop calculation, identified with the 4d string coupling $g = e^{\phi_4}$. These types of loop corrections violate the tree-level no-scale symmetry that appears at tree level. Recall that in a no-scale model, the gravitino mass behaves as $m_{3/2}^2\ \sim \ 1/u^r$,  and $e^{2\phi_4}\sim  1/\sqrt{su^r}$, where $u_i$ denotes a no-scale saxion and $r$ is a positive constant, so that the potential exhibits a runaway behavior at large moduli, as expected.
 On the other hand, we argue that the one-loop potential vanishes at small moduli, in particular at the so-called
 \textit{desert points} \cite{Long:2021jlv,vandeHeisteeg:2022btw}
 which maximise the value of the species scale.  In between, there is a dS hill with a maximum.
This structure results in the fixing of moduli at \textit{desert points}. This phenomenon is generic for any
no-scale moduli, and is due to the combined effect of two Swampland facts: \textbf{i)} the existence of a tower of states associated to 
each large modulus, and \textbf{ii)}  the fact that the cut-off of the theory, the species scale,  is moduli dependent.
 We can call this phenomenon {\it moduli self-fixing} since any no-scale modulus gets it just from its own tower of massive states.

This one-loop potential may have an important impact on the issue of moduli fixing. 
We have not yet done a thorough analysis, but e.g., in principle 
 there is no need to fix the no-scale K\"ahler moduli in Type IIB
through a non-perturbative superpotential, the mentioned one-loop 
self-induced potential is always there and fixes the moduli at values  ${\cal O}(1)$.
At those small values, in principle, we lose perturbative control of the theory, but we argue that the presence of those minima is robust.
Furthermore, we argue that the duality symmetries, inherent to string  theory,  may help us to get certain 
information about that limit in some simple cases.
Another worry is that such small values for the K\"ahler moduli in type IIB could be in conflict 
with the observed perturbative values of gauge couplings in the Standard Model (SM).
In type IIB, the SM gauge couplings constants are dictated by the vevs of the K\"ahler moduli and  (complex dilaton).
This worry is however not justified since the SM gauge couplings may have non-asymptotically free 
running, which would render the SM couplings perturbative in the IR. In fact, in specific MSSM-like string constructions, this non-asymptotic freedom is 
most often present.

The way we proceed in this paper is as follows. After a short review about towers and field dependent species scale,
we move in section \ref{section3}  to
 compute the one-loop potential for a general no-scale modulus (modulus contributing to SUSY breaking
in a non-SUSY Minkowski vacuum) with large vev. We take into account contributions coming both from the IR sector of states with a mass  $\simeq m_{3/2}$  and from the
tower of massive states. We show that the IR contribution is always negative, due to the hyperbolic structure of the Kähler metric of moduli in string theory, and
proportional to the cut-off $\Lambda^2$. On the other hand, the contribution from the tower is positive, independent of the cutoff and of order $e^{2\phi_4}\,m_{3/2}^2M_{\rm P}^2$.
This structure for the potential sends the modulus dynamically towards  ${ \cal O}(1)$  values, due to the field dependence of $\Lambda$.
  In section \ref{section4}, we make a second, more global computation,
again considering the sum over massive states, but expressed in terms of the species scale, obtaining an expression of the form
\beq
\delta V_{1-\rm{loop}} \ \simeq \ e^{2\phi_4}\,m_{3/2}^2 M_{\rm P}^2  \ g^{i{\overline i}} \frac {(\partial_i \Lambda ) } {\Lambda}  \frac {(\partial_{\overline i} \Lambda)}  {\Lambda }   \ .
\label{poten0}
\eeq
This reproduces the same qualitative results as the previous local computation, but it shows the presence of minima in Minkowski at the \textit{desert points},
which are those points in moduli space in which the species scale reaches its maximum. 
We observe that, when applied to Type IIB, the large-volume behavior of this one-loop correction scales as $\sim |W_0|^2/{\cal V}^3$, yielding a much more suppressed contribution compared to the tree-level term. Moreover, we argue that these corrections are consistent with the explicit string theory one-loop computations presented in Ref.~\cite{Berg:2014ama,Haack:2018ufg}, thereby providing a non-trivial check that strengthens our results.

In Section~\ref{section5}, we study string theory models, including cases such as a two-torus (analogous to the Enriques Calabi--Yau), for which the behaviour of the species scale has been specifically computed from topological string theory. The resulting structure is consistent with the findings of the previous section. In this case, the quantity $V/m_{3/2}^2$ is a modular invariant function that vanishes at the SL$(2,\mathbb{Z})$ self-dual points.

In section \ref{section6} we study how the presence of these one-loop potentials
 could affect the moduli fixing program, with a simple option being to replace non-perturbative superpotentials by 
the one-loop one. We work out in some detail moduli fixing in a  $(\mathbb{Z}_2\times \mathbb{Z}_2)$  orientifold, combining fluxes with the one-loop potential. 
One can, in principle, fix all untwisted moduli by combining fluxes and the one-loop potential, 
although, as usual,
RR tadpole cancellation is problematic. In  Appendix  \ref{apB}, we present a specific model with D6-branes, fluxes, and an MSSM-like spectrum
in which the tadpole issue can be substantially improved. In section \ref{section7}, we briefly discuss how this kind of one-loop potential represents a
specific example of IR-UV correlation. We also discuss its possible role in inflation. Section \ref{conclus} is left for some general discussion.

\section{Towers and the field dependent species scale}
\label{section2}

Let us just give a few definitions that will be used later in the paper.  One of the crucial properties that we have learnt in the context of the Swampland 
program \cite{Vafa:2005ui} in (QG) in recent years is the Swampland Distance Conjecture (SDC) \cite{Ooguri:2006in}. It states that when a modulus takes trans-Planckian values, a tower
of states (either KK-like or stringy \cite{Lee:2019apr,Lee:2019wij}) should become light. This has been tested to be correct in a large number of string vacua, see references in 
\cite{Brennan:2017rbf,Palti:2019pca,vanBeest:2021lhn,Grana:2021zvf,Harlow:2022ich,Agmon:2022thq,VanRiet:2023pnx} for detailed reviews, and it has also been motivated from bottom-up arguments \cite{Calderon-Infante:2023ler,Aoufia:2024awo}. 
The structure of towers in string theory takes the following general form for the mass of the $n$-th level state
(see e.g.\cite{Castellano:2021mmx,Castellano:2022bvr})
\beq
m_n^2 \ =\ n^{2/p} m_0^2 \ ,\ n=0,...,N
\label {torre}
\eeq
Thus, e.g., for a single KK tower one has $p=1$ and the familiar relation $m_n^2=n^2m_0^2$ follows. The case $p=2$ would correspond e.g., to an effective tower 
for two KK towers.  Also the {\it gonion} towers in \cite{Casas:2024ttx,Casas:2024clw}
are $p=2$ towers.   On the other hand, we have also learnt that the fundamental scale of the theory is not the
Planck scale but rather the Species Scale $\Lambda $
\cite{Dvali:2007hz,Dvali:2007wp,Dvali:2008ec}, which can be much lower depending on the
number of states   $N$ below $\Lambda$. 
The Species Scale in d-dimensions is related to
the number of species in a tower  $N$ by
\beq 
\Lambda ^{d-2}\ \simeq \ \frac {M_{\rm P}^{d-2}} {N} \ .
\label{species}
\eeq
 This is the value of the species scale at a given direction in the moduli space, in which a particular tower is coming down. 
 More generally, it is important to realise that the species scale is a {\it  function of all the moduli of the theory}
 \cite{vandeHeisteeg:2022btw,vandeHeisteeg:2023ubh,vandeHeisteeg:2023dlw,Castellano:2023aum,Aoufia:2025ppe},
 giving different results depending on the 
 moduli direction. Thus, it is also defined in the bulk of moduli space where there are no towers of states.  Note that, in a given direction in which a 
 tower of states arises, the value of the species scale is related to the lightest state in the tower. Indeed, for the heaviest state in the 
 tower one has
 \beq
  m_N^2\ =\ \Lambda^2=N^{2/p}m_0^2 .
  \eeq
Combining this with eq.(\ref{species}) one gets the relationship
\beq
\Lambda \ =\ m_0^{p/(d-2+p)}
\label{torrespecies}
\eeq
in Planck units. This applies to any of the many towers that a string vacuum may have, with the species scale function $\Lambda$ taking values along the
different directions.  
Concerning string towers,  although one would also naively say that one has  $p=2$, it turns out that a more appropriate
description is $p=\infty$ 
since the degeneracy at each level grows like $e^{\sqrt{n}}$
\cite{Castellano:2021mmx,Castellano:2022bvr}.  In particular note that for $p\rightarrow \infty$ 
eq.(\ref{torrespecies}) yields $\Lambda \rightarrow m_0^2=M_s^2$ with $M_s=m_0$ the string scale, which is correct since the degeneracy at each 
level is so large that the species scale coincides with the string scale.

The species scale depends on all the moduli of the theory and carries information on the behaviour of the fundamental QG cut-off in all possible
moduli directions, including the infinite singular limits mentioned above, but it also carries information about the bulk of the theory in which no
large towers emerge. So given a string vacuum, it would be very informative to have a closed expression for $\Lambda$ as a function of the moduli. 
It has been recently realised \cite{vandeHeisteeg:2022btw,vandeHeisteeg:2023ubh,vandeHeisteeg:2023dlw,Castellano:2023aum}
that in string vacua with enough SUSY generators one can obtain in some cases explicit expressions for the species scale. 
In particular one can consider BPS protected higher derivative operators of the general form
\beq
S_{\rm BPS} \ =\ \frac {1}{2\kappa _d^2}
\int d^d x \sqrt{-g} N {\cal F}_n^{(d)} \frac {{\cal R}^{2n} }{M_{p;d}^{4n-2}}\ ,
\eeq
with $n=2$ in the case of 32 and 16 supercharges, and $n=1$ for the case of 8 supercharges,
like 4d ${\cal N}=2$ CY compactifications of Type II string theory.  One expects the mass scale
controlling these operators to be given by the cut-off of the theory so that one has
\beq
\Lambda  \ \simeq \ M_{{\rm p}}^{(d)}\, {\cal F}_n^{-\frac {1}{4n-2}} \ .
\eeq 
Thus one can get a field dependent expression over all moduli space if one is able to compute the Wilson
coefficients  ${\cal F}_n^{(d)}$. You can do this, as we said, when the operator has some BPS protection. 
Of more direct interest to us here will be the case of 4d ${\cal N}=2$ theories obtained from  10d Type II 
compactifications in a CY manifold, in which case one has
\beq
S_{{\cal N}=2} \ =\ \frac {1}{\kappa_p} \int  d^4x \ N\ 
\frac {({\cal R}\wedge {\cal R})}{M_{\rm P}^ 2} \ ,
\eeq
with $N=\Lambda^{-2}$.  Here $N$ will be a function in Type IIA of the $h_{11}$ vector moduli and $(h_{21}+1)$ hypermultiplets 
(the opposite in Type IIB).  It turns out that the dependence on vector moduli may be explicitly computed in terms of topological string theory \cite{Bershadsky:1993ta}.
In particular it was argued in  \cite{vandeHeisteeg:2022btw,vandeHeisteeg:2023ubh}
that  $N$ may be obtained from the genus-one free energy of topological strings propagating in the CY.
Thus in a class of theories the species scale (or at least its dependence on vector moduli) may be computed exactly, including perturbative and non-perturbative effects.
An example of this is \cite{Bershadsky:1993ta,vandeHeisteeg:2022btw,vandeHeisteeg:2023ubh}
the species scale for theories including a torus, like the Enriques CY which is $K3\times T^2/{\mathbb{Z}_2}$, see
eq.(\ref{N-toro})  in section (\ref{section5}).
One important general property of the species scale is that it contains all duality invariances in the theory.  An example of this is the
SL$(2,\mathbb{Z})$ modular invariance associated to the $T^2$ in those examples. The associated species scale is modular invariant. This 
is also known to happen in higher-dimensional string vacua like maximal sugra toroidally compactified 
\cite{Green:1997tv,Green:2010kv,Green:2010wi}.
This modular invariance will be an important guide in order to try to obtain information on the form of the effective potential in the following sections.

Eventually will be interested in ${\cal N}=1 $, 4d theories, which have less than the required SUSY in the above computations. However, in many cases 
the compactifications are an orientifold reduction of an ${\cal N}=2$ theory and one expects the essential features of the species scale 
to be a projection of the results with more SUSY.

\section{Scalar potential from a tower}
\label{section3}

In this section  we will study the scalar potential induced at one-loop in 4d theories with spontaneously 
broken ${\cal N}=1$ supersymmetry within the general context of string vacua.
Since one of the relevant applications is the issue of moduli fixing, we will have in mind in particular 
CY orientifold theories in Type II string theory, although we believe the results here will be quite 
general and will apply to other corners of string vacua.
Thus
our starting point will be classical non-SUSY vacua in Minkowski space in 4D, obtained as no-scale solutions. The prototypical examples
are GKP-like configurations in Type IIB theory \cite{Giddings:2001yu}, although we will also consider some slight variations.  We will mostly work 
within the context of Type IIA orientifold theories, mostly because of the examples in
section (\ref{section6}) are more conveniently expressed in that setting, but the results and ideas may be easily translated to the mirror Type IIB language.
These kinds of theories have chiral multiplets $Y_a$ which explicitly appear in the flux superpotential $W_{\rm flux}(Y_a)$, with $D^aW_{\rm flux}=0$.
The rest of the moduli denoted $Z_i$ may contribute to the SUSY-breaking and their auxiliary fields verify a no-scale condition
$g_{i{\bar i}}|F^i|^2=3m_{3/2}^2$ leading to Minkowski. The $z_i$ remain at this level as classical moduli, and we will call them {\it no-scale moduli}.
Accordingly, in the Type IIB GKP case, the $Y_a(Z_i)$ will be complex structure + complex dilaton (K\"ahler moduli) respectively.

The focus will be on studying the one-loop contributions to the vacuum energy as a function of the no-scale moduli, $V_{1-\rm{loop}}(Z_i,{\bar Z}_{\bar i})$.
We will denote for the axions and saxions $Z_i=\eta_i+iz_i$, $Y_a=v_a+iy_a$. 
In our initial EFT  computation in this section we will consider the limit
in which a single no-scale saxion goes to infinity, $z_i\rightarrow \infty$, and will compute the 
one-loop scalar potential close to that limit.
The swampland distance conjecture tells us that a tower of states will become 
massless in this limit. Thus, in our loop computation, we should  include both zero modes as well as the massive modes in the corresponding tower.
Crucially, we will take as the cut-off in the computation the scale at which strong quantum gravity effects cannot be neglected,
the species scale $\Lambda $. 
 This is reasonable since we know that string theory is UV finite and any divergence (at the level of the EFT) should be cut-off by the fundamental scale of the theory, which is the species scale. Of course, a full string theory computation includes infinite sums and still is UV finite  with no need for any explicit UV cut-off. 
It is when going to an EFT below the fundamental scale that an effective cut-off, the 
species scale  is required to describe the theory, including loops.
Thus, one should include in the EFT all of the states which lie
below the species scale and inserts them, e.g., in loop computations.
 Note that the number of such states is necessarily finite, e.g., for fixed radius $R$ there is room only
for a finite number of states $N=m_{\rm KK}R$ below the species scale. From the point of view of the EFT, we believe it is not correct to sum over an infinite number of states,
 including states above the species scale.
Thus, e.g., going above the string scale we would rather be led to an equivalent dual theory with windings replacing momenta.  
This is, e.g., the approach followed in the 
\textit{emergence} computations in which it is shown how loop corrections can generate the kinetic terms in the theory
\cite{Heidenreich:2017sim,Grimm:2018ohb,Heidenreich:2018kpg,Castellano:2022bvr,Marchesano:2022axe,Blumenhagen:2023yws,Blumenhagen:2023tev,Blumenhagen:2023xmk,Casas:2024ttx,Blumenhagen:2025zgf}. The 'weak' version of this proposal has been tested in many string theory vacua, and the successful results
require summing over all terms in a tower below the species scale
\footnote{Although we will perform sums over towers of massive modes, similar to the ones performed in the emergence approach,
we will not make use of any emergence argument in our discussion.}.  As a final comment in this regard, we would like to emphasize that the usual Casimir contributions computed in field theory are obtained by summing over an infinite tower of modes and subsequently performing a regularization procedure. We want to emphasize in this work that, in the context of quantum gravity 
in which a natural cut-off appears, such divergences are naturally avoided once the fundamental cut-off, namely, the species scale, is introduced, rendering all contributions finite. This mimics what happens in full string theory, where no UV divergences arise.

There is the question of whether 
one should use instead as a cut-off the lightest mass of a tower of species like e.g., the KK scale $M_{\rm KK}$,
see e.g. \cite{Cicoli:2013swa,Burgess:2023pnk,ValeixoBento:2020gpv,Blumenhagen:2023yws,Calderon-Infante:2025ldq}
 for some discussions on this issue. 
Indeed for generic correlators in the EFT 
the appropriate cut-off would be $M_{\rm KK}$. However, if we are interested in operators that arise after dimensional reduction of relevant or marginal operators in the higher-dimensional theory, one should sum over the finite
number of states below the species scale. In particular, the one-loop vacuum energy computation at hand in this section is just a trace over states
and is not sensitive to details of the EFT like higher dimensional operators, depends  only 
on the spectrum (light and heavy) of the theory and not on the specific form of the EFT.  It would be inconsistent to neglect the
contribution of the finite number of states, however small, below the Species Scale.

As we said, the species scale $\Lambda$ depends on all moduli, but we will take the $Y_a$ moduli fixed at their
classical vev and consider thus the species scale as a function of the $Z_i$, $\Lambda (Z_i,{\bar Z}_{\bar i})$. Since the cut-off will 
appear explicitly in the final result for the one-loop computation, its field dependence plays an important role in the dynamics.

Let us start by recalling the expression for the one-loop contribution to the effective potential in (broken) supersymmetric theories. For completeness, we present this computation in Appendix \ref{appendix one loop}.
It is  given at leading order by \cite{PhysRevD.7.2887,PhysRevD.7.1888,Schwartz_2013}
\begin{equation}
    V_{\rm total}^{\rm eff} = \frac{1}{(8\pi)^2}\left(-\text{Str}\mathcal{M}^0\Lambda^4 + 2\,{\rm Str}\mathcal{M}^2\Lambda^2 - 2\, {\rm Str}\mathcal{M}^4 \log\left(\frac{\Lambda}{\mathcal{M}}\right)\right) \ ,
    \label{potencial}
\end{equation}
where we have defined the supertraces
\begin{equation}
    {\rm Str}\mathcal{M}^a = \sum_n (-1)^{2j_n}(2j_n + 1)( m_n)^a,\quad a=0,2,4,
    \label{supertraza}
\end{equation}
and $j_n$ is the spin of each particle.  Note that in a SUSY theory one has Str$\mathcal{M}^0=0$, so that the leading term will be the second one.
The following term in ${\rm Str}{\cal M}^4$ may be shown to be subleading, see Appendix \ref{app: strm4}.
 \begin{figure}[h!]
            \hspace{2em}
            \includegraphics[width=0.9\linewidth]{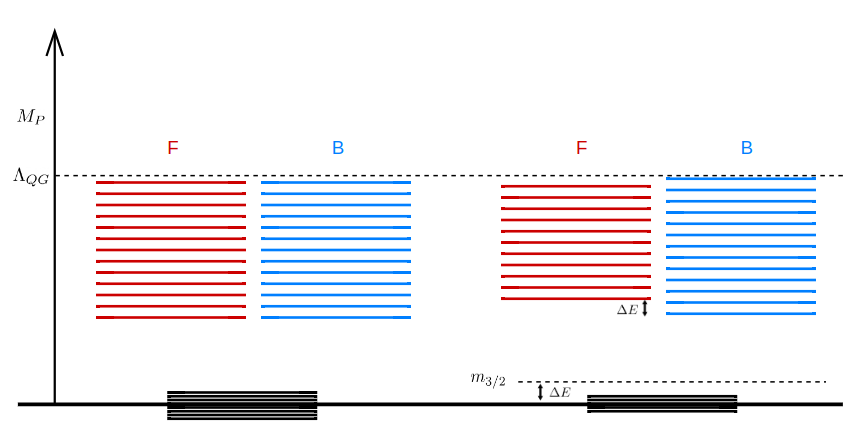}
            \caption{The different towers of states and the light states with unbroken SUSY are depicted on the left of the picture. On the right, SUSY is spontaneously broken, the gap between states is given by the gravitino mass, and there are generally fewer light states in the theory.}
            \label{fig:structure scales}
        \end{figure}
Let us start by computing the contribution of the light modes. In a theory with F-term SUSY breaking there is a general expression for the supertrace 
for a ${\cal N}=1$ supergravity theory with $N_0$ chiral multiplets,  see e.g. the paper  by Ferrara and van Proyen
\cite{Ferrara:2016ntj}:
\beq
{\rm Str}{\cal M}^2\ =\  2(N_0-1)m_{3/2}^2 \ +\ 2e^{K}{\cal R}_{i{\bar j}}(D^iW)({\overline D}^{\bar j}{\overline W}) \ .
\label{ferrara}
\eeq
Here $N_0$ is the number of chiral multiples, and ${\cal R}_{i{\bar j}}$ is the Ricci tensor in moduli space. 
Also  \( D^i W = g^{i \bar{j}} \nabla_{\bar{j}} \overline{W} \) denotes the K\"ahler-covariant derivative of the superpotential, and  $m_{3/2}$ is the gravitino mass.
Crucially,  our chiral multiplets here will be  string moduli, which typically have  hyperbolic metrics at large values like
\beq
g_{i{\bar i} }\ =\ \frac {2}{a_i^2} \frac {1}{ |z_i - {\bar z}_{\bar i}|^2} \ .
\eeq
Where $a_i$ depends on the explicit example.  Then the Ricci tensor has the form \footnote{ We are not considering directions in moduli space like those studied in \cite{Marchesano:2023thx,Marchesano:2024tod,Castellano:2024gwi,Blanco:2025qom} in which an
additional positive piece may appear in the right-hand side corresponding to systems decoupling from gravity.}
\beq
R_{i{\bar i}}\ =\ - \ g_{i{\bar i}} (N_0\ +\ 1) \ . 
\eeq
Thus, inserting this expression in \eqref{ferrara}, one obtains for the rightmost piece
\beq
2e^{K}{\cal R}_{i{\bar i}}(D^iW)({\overline D}^{\bar i}{\overline W})\ = -\ 2e^K (N_0+1) g_{i{\bar i}}
(D^iW)({\overline D}^{\bar {\bar i}}{\overline W}) \ = \ -6m_{3/2}^2 (N_0+1) \ ,
\eeq
where we have imposed the no-scale condition in the last step.  Thus one obtains in this case
\beq
{\rm Str}{\cal M}_{{\rm light}}^2\ =\ -4 m_{3/2}^2\,
 \left( N_0  + 2\right) \ .
 \eeq
Note that we have arrived at a negative definite term, which is important in what follows.
This is a direct consequence of the fact that in string compactifications the moduli metrics are hyperbolic.
In particular,  if we had in the massless sectors a sufficiently large number of chiral fields
with canonical metrics the supertrace could be positive \footnote{ This is in agreement with the {\it light fermion 
conjecture} of ref.\cite{Gonzalo:2021fma} which requires $Str({\cal M}^2)< 0$ for Minkowski vacua.}.

\begin{figure}
  
\hspace{2em}
\includegraphics[width=1.0\linewidth]{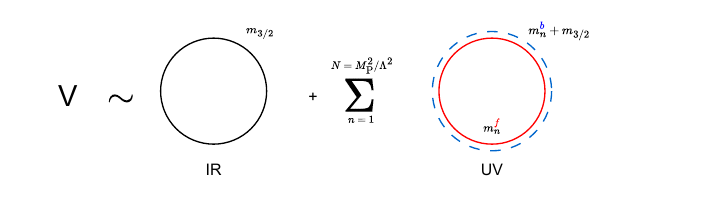}
    \caption{Schematic representation of the one-loop computation described in the text. The first loop corresponds to summing over light states, while the second loop corresponds to summing over the associated tower of states with masses below the cutoff. }
    \label{fig:one loop UV+IR}
\end{figure}

Consider now the contribution of the states in the tower.  They have canonical kinetic terms so that
 the bosons will get a shifted mass squared compared to fermions  at the level $n$,
\beq
m_{b,n}^2\ -\ m_{f,n}^2  \ =\ m_{3/2}^2 \ .
\eeq
This shift in the boson masses is well known to happen in spontaneously broken  ${\cal N}=1$ supergravity for chiral multiplets with 
canonical kinetic terms (see.e.g. \cite{Hall:1983iz} and references therein).
One can then compute the total contribution by summing over the tower of massive states with $\simeq N$ levels
\footnote{ Since there are at least two chiral multiplets at each level $n$ we included an overall factor of 4, but in fact 
we are neglecting here the detailed degeneracy at each level, except for the $e^{\sqrt{n}}$ Hagedorn degeneracy in the
case of string towers.}.
\begin{equation}
    {\rm Str}\mathcal{M}^2\Lambda^2 \simeq  4\Lambda^2\sum_n^N( n^{2/p} m^2 + m_{3/2}^2 - n^{2/p} m^2 ) \simeq  4\Lambda^2 N m_{3/2}^2 \simeq 4m_{3/2}^2 M_{\rm P}^2 \ ,
\end{equation}\newline
where $m$ is the mass of the lightest state in the tower and $p$ is the density parameter described in section \ref{section2},
and also we have used the defining equation $\Lambda^2\simeq M_{\rm P}^2/N$.  Interestingly, the result is independent of the
UV cut-off $\Lambda$. It is also independent of the density parameter $p$ of the tower. Thus, it applies not only for KK towers but also for the string case $p=\infty$. The reader may see a similar sum computation taking into account the Hagedorn degeneracy
at each level in section $(3.1.2)$ of ref.\cite{Castellano:2022bvr} and in Appendix \ref{appendix one loop}.

\begin{figure}[h!]
            \centering
            \includegraphics[width=0.5\linewidth]{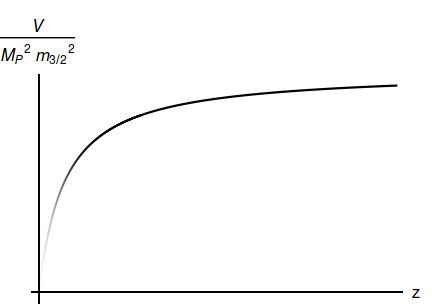}
            \caption{Qualitative behaviour of equation \eqref{V1loop} in terms of a single scalar field. The figure shows how, at the field theory level, the one-loop potential contribution drives the scalar field to small vevs. }
            \label{fig:single field}
        \end{figure}

In summary, for large $N$ the total potential has the structure
\beq 
V_{1-\rm{loop}} \  \simeq  \ \frac {m_{3/2}^2M_{\rm P}^2}{(8\pi)^2}   \left( c \ - \ \frac {\eta}{N(z_i,{\bar z}_{\bar i}) }\right)  \  ; \ \ \eta = 8\left(N_0+2\right)\ ,
\label{V1loop}
\eeq
with $c$ a positive constant of order one. The first term originates from the tower, while the second arises from the light modes (including the gravitino). 
Note that for fixed gravitino mass the potential is minimised as $N\rightarrow {\cal O}(1)$, which in terms
of moduli corresponds to $z_i\sim {\cal O}(1)$, with the species scale $\Lambda$ approaching its 
maximum value close (below) $M_{\rm P}$.
Thus the potential tends to send the modulus to a small value, see Fig. 
 \ref{fig:single field}.  Note that, as we said,  for this to be the case it is crucial to have a negative 
sign in the zero mode contribution,  which in turn rests on the hyperbolic character of the metric.  This will apply generically to all
no-scale moduli $z_i$ contributing to SUSY-breaking; whenever they grow, they themselves create a potential which drives them to 
$z_i\sim {\cal O}(1)$. This is why we call this mechanism {\it moduli self-fixing}.

Consider the simplest example of a CY with $h_{11}=1$ in Type IIA string theory. The K\"ahler potential for the Kähler moduli at large
volume has the standard no-scale form
\beq 
K(T,T^*) \ =\ - 3\log (-i(T-T^*)) \ .
\eeq
We know that at large $t={\rm Im}T$, a tower of $D0$ branes will appear.  The modulus $t$ is massless at tree level, but at one-loop
a potential appears in the form
\beq 
V_{1-\rm{loop}} \  \simeq  \ \frac { g^2 m_{3/2}^2M_{\rm P}^2}{(8\pi)^2}\left( c \ - \ \frac {12}{N(T,{\bar T}) }\right)\ .
\eeq
 It is known that the species number $N$ in type IIA CY compactifications is linear in $t$  for large modulus.
  So that $N\rightarrow t$ and the potential decreases as $t$ decreases,  and 
  the modulus $t$ is driven to small values. We will argue in the next section that in fact the potential vanishes
  and reaches a minimum at the point where $N$ is minimal.  It should also be noted that, when embedding this analysis in string theory, one must incorporate the loop-counting coupling parameter, which in four dimensions is given by $g = e^{\phi_4}$. In the following analysis, we will incorporate this coupling in all computations.

\begin{figure}[h!]
            \hspace{9.5em}
            \includegraphics[width=0.65\linewidth]{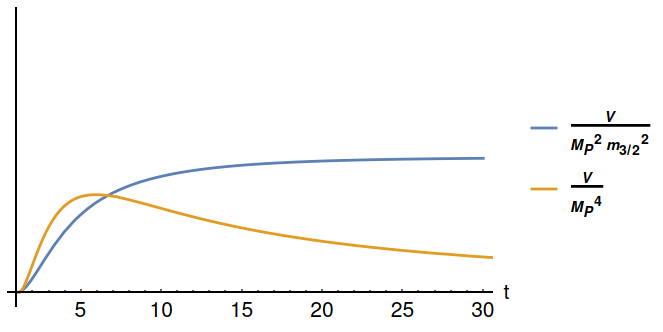}
            \caption{The blue scalar potential corresponds to the one-loop contribution with a fixed gravitino mass. The orange potential corresponds to a modulus-dependent gravitino mass, as observed in simple flux compactification settings. In this latter case the potential vanishes at the {\it desert points} and at infinity, performing generically dS maxima in between. In this simple one modulus toy model we have fixed the $e^{\phi_4}$ to a constant.
         }
            \label{fig:potentials}
        \end{figure}

  In fact the gravitino mass is also modulus dependent and gives an additional factor.
  Thus consider the simplest example with an RR flux 
  $F_6=e_0$ in IIA. Then the gravitino mass in these examples is $m_{3/2}^2=e_0^2/(8t^3)$. This factor modifies 
  the form of the potential so that for large $t$ it becomes run-away with $V\simeq  g^2\,e_0^2/t^3$. 
  Still,  the minimum at small $ t \simeq {\mathcal O}(1)$ remains, with a dS maximum in between both limits.
  The general qualitative shape is as in the orange line in Fig. \ref{fig:potentials}.  Thus the claim in this example is that, after SUSY-breaking,
  the tower of $D0$'s in IIA induces a potential for the volume K\"ahler modulus of the said qualitative form.
  This looks like a generic pattern for any no-scale modulus breaking SUSY.
 
Recently, in \cite{ValeixoBento:2025yhz}, the authors studied  5- and 4-dimensional M-th dS maxima with the inclusion of Casimir energies.  In their case it is the
first contribution of eq.(\ref{potencial}) which causes the positive contribution to the potential. It is  made  non-vanishing  by playing with the fermion boundary conditions.
In our case, on the contrary,  it is the second term in eq.(\ref{potencial}) which plays the leading role.

\section{The scalar potential and the Species Scale}\label{section4}

The previous computation concerned a single asymptotic direction for a no-scale modulus. Although we found 
a structure leading to minima at small moduli, with a dS maximum and an asymptotic runaway behaviour, we
could not say much about what the behaviour of the potential could be at small moduli.
We are now going to discuss an alternative computation of the one-loop potential which will
allow us  to extrapolate the asymptotic results into the bulk of moduli space in a 4d theory 
with spontaneously broken ${\cal N}=1$ SUSY.

One can obtain  a more useful  expression  for the one-loop potential 
by computing one-loop corrections to the \( \mathcal{N}=1 \) superfield correlator for the modulus (specifically, the D-term correlator \( \langle  Z_i{\bar Z}_{\bar i} \rangle_D \)). Here, \( Z_i\) is the \( \mathcal{N}=1 \) superfield defined by
\begin{equation}
 Z_i =    (\eta_i+iz_i) + \theta \tilde{z_i} + \theta^2 F_i,
\end{equation}
where \( \theta \) is the Grassmann coordinate, and \( F_i \) is the supergravity auxiliary field associated with the modulus  superfield \( Z_i \).
Setting $Z_i$ to its bosonic component \( Z_i = \eta_i+ iz_i\) one can see one recovers a correction to the K\"ahler metric of the modulus, \( g_{i\bar{i}} \), as computed in references \cite{Grimm:2018ohb,Heidenreich:2017sim,Heidenreich:2018kpg,Castellano:2022bvr,Castellano:2023qhp,Blumenhagen:2023yws,Casas:2024ttx}. However, evaluating the same one-loop diagram with \(  Z_i = \theta^2 F_i \) instead leads to a contribution to the vacuum energy proportional to \( |F_i|^2 \).
\begin{figure}[h!]
    \hspace{4.em}
    \includegraphics[width=0.8\linewidth]{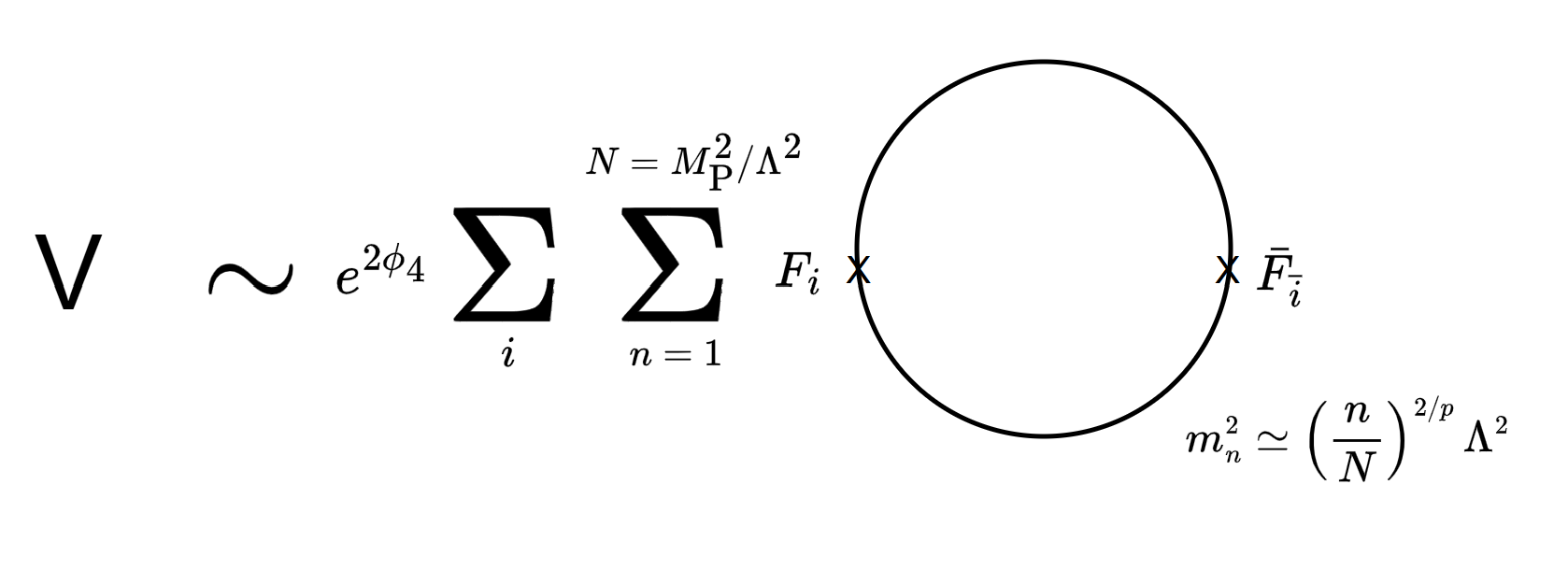}
    \caption{Schematic representation of the one-loop computation carried out from the F-term auxiliary 
    field insertion in supergravity. The loop corresponds to summing over the tower of states, and the sum in $i$ corresponds to summing over the no-scale moduli  contributing to SUSY breaking. The 4d dilaton $e^{\phi_4}$ factor represents the loop-counting coupling parameter $g$.}
    \label{fig:FFpotential}
\end{figure}
To be more precise, let us briefly recall how loop corrections to the metric arise in theories involving a modulus scalar, as described in \cite{Castellano:2022bvr}. For clarity, consider the case of a single Kaluza-Klein (KK) tower, although, as shown in \cite{Castellano:2022bvr}, the results extend to arbitrary values of the density parameter \( p \), including the string-theoretic case.
\newline
  The modulus $z_i$ has {\it Yukawa couplings}  to the towers of the form
$\partial_im_{ni} = n\partial_im_{0i}$, with $m_{0i}$ the KK scale.  Then a one-loop graph would give rise to a correction  \footnote{ In fact for $d=4$ at each level there is a factor  $\log(\Lambda/m_{ni})$ multiplying. 
However one can see that this log disappears once one sums over the whole tower, see ref.\cite{Castellano:2022bvr}.
We are also ignoring numerical factors of order one. } to the metric \cite{Castellano:2022bvr}
\begin{align}
\delta g_{i{\overline i}}  \ \simeq \ \frac {4\,g^2}{(8\pi)^2} \sum_n  n^2  (\partial_im_{0i})(\partial_{\bar i}m_{0i}) \ 
\simeq  &\frac {4\,g^2}{(8\pi)^2} N^3  \ (\partial_im_{0i})(\partial_{\bar i}m_{0i})\label{correctiondG}\\
&\simeq  \  \frac {4\,g^2}{(8\pi)^2}  \frac {(\partial_im_{0i})(\partial_{\bar i}m_{0i}) } {m_{0i}^2}\,M_{\rm P}^2\ \notag,
\end{align}
where the coupling parameter $g$ in the loop calculation is given by $g=e^{\phi_4}$ and we have approximated the sum by an integral, and in the last step used $\Lambda=Nm_{0i}$ and $M_{\rm P}^2= N\Lambda^2$. This structure is the same for all $p$. 
Given a tower saturating the species scale one can use $m_{0i}= \Lambda^{(d-2+p)/p}$ as discussed in section \ref{section2}, and write \eqref{correctiondG} in terms of the species scale \footnote{Substituting in \eqref{correctiondG} the species scale in terms of the mass of the tower is not an innocuous replacement since the species scale is sensitive to IR physics and
extends over all moduli space. For example, it could reveal modular symmetries that the field possesses and not just its asymptotic behaviour, see the discussion below. }
\beq
\delta g_{i{\overline i}} \ \simeq \ \frac {4\,g^2 }{(8\pi)^2}\frac {(\partial_i\Lambda )(\partial_{\bar i}\Lambda)} {\Lambda^2}\,M_{\rm P}^2\ \ .
\label{metrica}
\eeq
In superspace notation, the correction to the kinetic term may be written as a D-term,
\beq
\delta{\cal L}_{kin} \ =\ \left[ \delta g_{i{\overline i}} Z^i   {\overline  Z}^{\overline i} \right]_D\ \simeq
 \  \left[ \frac {4\,g^2 }{(8\pi)^2} \  \frac {(\partial_i\Lambda)(\partial_{\bar i}\Lambda)} {\Lambda^2}\ 
\   Z^i {\overline  Z}^{\overline i}\right]_D\, M_{\rm P}^2 .
\eeq
From $(Z^i {\overline  Z}^{\overline i})_D$   one recovers the one-loop correction to the modulus
kinetic term metric. However, setting   $Z^i = \theta^2F^i$ one recovers a contribution to the vacuum energy to leading order 
\footnote{Note that the potential vanishes for SUSY AdS vacua in which $V=-3m_{3/2}^2M_{\rm P}^2$.}
\begin{equation}
   \delta V\ \simeq \  \frac {4\,g^2\,m_{3/2}^2M_{\rm P}^2 }{(8\pi)^2}  \left(\frac{V}{m_{3/2}^2}+3\right) \,g^{i\bar{i}}\,\frac{\partial_i\Lambda\partial_{\bar{i}}\Lambda}{\Lambda^2}\, 
   \label{conV}
\end{equation}
where we have used that $F_iF^{\bar{i}}=V+3\,m_{3/2}^2$ and it holds for  general $V$. Here $g^{i\bar{i}}$ is the tree level metric.
Insisting though, as in the previous section, on no-scale fields with Minkowski vacuum, one obtains,
\beq
\delta V \ \simeq \ \frac {4\,g^2\,m_{3/2}^2 M_{\rm P}^2 }{(8\pi)^2}  \ g^{i{\overline i}} \frac {(\partial_i \Lambda ) } {\Lambda}  \frac {(\partial_{\overline i} \Lambda)}  {\Lambda }   \ .
\label{poten1}
\eeq
One can also write
\beq
\delta V \ \simeq \ \frac {g^2\,m_{3/2}^2 M_{\rm P}^2}{(8\pi)^2}  \ g^{i{\overline i}} \frac {(\partial_i N ) } {N}  \frac {(\partial_{\overline i} N)}  {N}   \ .
\label{poten2} 
\eeq
It is interesting to note that the one-loop correction to the potential can be expressed as
\begin{equation}
\delta V \simeq \frac {4\,g^2\,m_{3/2}^2 M_{\rm P}^2}{(8\pi)^2}  \, (\lambda_{\text{sp}}^{i})^2 ,
\end{equation}
where  $\lambda_{\rm sp}^i$ characterizes the variation rate of the species scale:
\begin{equation}
\lambda_{\text{sp}}^i = \left| \frac{\nabla_i \Lambda}{\Lambda} \right| .
\end{equation}
The asymptotic form of this parameter  has been analysed in the context of the Swampland program  and  is known to be constrained by the conditions
\cite{Calderon-Infante:2023ler,Etheredge:2023usk,vandeHeisteeg:2023ubh}

\begin{equation}
   \frac {1} {d-2} \geq \lambda_{\text{sp}}^2  \ \geq \ \frac{1}{(d-1)(d-2)} ,
\end{equation}
where $d$ is the number of dimensions.
The upper (lower) bounds are in fact saturated for an asymptotic string (KK) limit respectively.
Thus, in the large-modulus regime, the potential asymptotically approaches a value independent of the UV cutoff:
\begin{equation}
\delta V \simeq g^2 m_{3/2}^2 M_{\rm P}^2 ,
\end{equation}
in agreement with the result obtained in section (\ref{section3})  in which a leading constant term $\sim m_{3/2}^2M_{\rm P}^2$
appeared (taking the coupling $g$ as a constant).

An important comment is in order.
This computation has been made for large $N$ (and moduli).
However we will make here the, admittedly, non trivial assumption that eq.(\ref{poten1}), derived in the large modulus limit, applies also in the bulk if one
uses a field dependent species scale  $\Lambda(Z_i, {\bar Z}_{\bar i})$ defined over all moduli space. 
 This is reasonable taking into account that   $\Lambda$  has in principle information about the bulk, and should provide 
 for a consistent extrapolation to points away from the singularities. 
 
 One can further  motivate this assumption based on 
 results from the case in which $\Lambda$ is a modular invariant function of a e.g., SL$(2,\mathbb{Z})$ 
 duality in which bulk physics and the asymptotic limits are 
 related by modular invariance. 
 Let us specify here a bit more about this, advancing some results described in detail in section \ref{section6}.
 There, we consider an example in which the species scale as a function of a modulus $U$ is known, and then the 
 potential,  as computed from  eq.(\ref{poten1}), is proportional to the modular invariant function
 $\sim u^2|{\tilde G}_2(U,U^*)|^2$, where ${\tilde G}_2$ is the (non-holomorphic ) weight-2 Eisenstein modular form. 
 Strictly speaking we only know eq.(\ref{poten1}) to be correct for large $u$, in which limit one has 
 $u^2|{\tilde G}_2|^2\rightarrow u^2$. Thus the idea is that modular invariance dictates that $u^2$ is
 completed in the bulk as  $u^2\rightarrow u^2|{\tilde G}_2|^2$.  But this corresponds to extending the 
validity of the computation eq.(\ref{poten1}) to the bulk in moduli space. Thus modular invariance justifies a posteriori the
extrapolation of eq.(\ref{poten1}) to the bulk.

 It is also important to  note  that,
  at this level,  this gives information on the scalar potential only around certain regions of moduli space, involving one single modulus.
The full potential will depend on all moduli in a given vacuum and will not be simply given by the sum of 
these individual pieces.  Still, this local behaviour of the potential is enough to discuss local minima.

There are a couple of important points concerning the potential in eq.(\ref{poten1}).  First, it is positive definite. Secondly, 
it has thus minima at points with $\partial_i\Lambda = 0$. This is interesting, because those points 
correspond to the {\it desert points} of the theory \cite{vandeHeisteeg:2022btw}. Thus we see that the dynamics reach Minkowski 
minima at the \textit{desert points} of the theory. In theories with duality symmetries, they correspond to self-dual points.
It is well known that the gradient of a modular invariant form vanishes at these points (see e.g.,
the appendix in \cite{Cvetic:1991qm} for some properties of modular forms).

 To see that the present computation also yields a negative correction going like  $\sim -1/N$ 
we have to be a bit more specific about the form of the species scale and the class of models we are discussing. 
To be more explicit, we will consider first the class of CY Type II compactifications leading to ${\cal N}=2$ theories in 4d. An additional
orientifold projection may lead to ${\cal N}=1$  (with some ${\cal N}=2$ subsector), which one can expect to inherit much of the results in the discussion.

\section{Moduli dependent Species Scales}\label{section5}

Let us review some facts about the species scale in ${\cal N}=2$ CY type II compactifications. As argued in \cite{vandeHeisteeg:2022btw,vandeHeisteeg:2023ubh} one can obtain the
piece of the species scale depending on vector multiplet moduli with the tools of Topological Field Theory. We are just going to present some general results 
obtained for $N(Z_i,{\bar Z}_{\bar i})= \Lambda^{-2}$ in \cite{Bershadsky:1993ta,vandeHeisteeg:2022btw,vandeHeisteeg:2023ubh}, in particular the large (vector multiplet) modulus limit of this function.
The details depend on the particular type of infinite limit, e.g., large volume, emergent string limit, or 6d decompactification, as well as the Hodge 
numbers of the CY compactification. For all relevant infinite distances, parametrizing the limit in terms of a single large modulus  $s$, one gets the
structure \cite{vandeHeisteeg:2023ubh}
\beq 
N\ =\ \frac {2\pi c_2}{12} s \ -\ \beta \log(s) 	 \ +\ \tilde{N_0} \ + \   {\cal O}(s^{-1}) \ .\label{expansion}
\eeq
Here, $c_2$ is the relevant part of the second Chern class of the CY. The value of $\beta$ depends on IR details,
in particular the Hodge numbers $h_{11},h_{21}$, and is in general a positive number, see ref.\cite{vandeHeisteeg:2023ubh}.
The constant $\tilde{N_0} $ denotes the number of states at the desert point.
Inserting \eqref{expansion} in \eqref{poten1}, up to subleading logarithmic terms 
\footnote{Such logarithmic terms are negligible as long as $\log(s)\lesssim {\tilde N}_0/\beta$, which allows for exponentiaaly 
large values of $s$, as one can check using the sample values of $\beta$ in ref.\cite{vandeHeisteeg:2023ubh}.}
takes the form,
              \beq
              V_{1-\rm{loop}}   \ \simeq \  \frac {g^2\,m_{3/2}^2 M_{\rm P}^2}{(8\pi)^2}
              \left( 1\ -\  \frac {\eta}{s}\right)^2 ,\ \quad \eta=\frac {12 \tilde{N_0} }{2\pi c_2 }.
              \label{poton}
              \eeq
              Thus, indeed, $V_{1-\rm{loop}}$ has the behaviour of a plateau and a negative one-loop correction of the form $\sim (-1)/N$, as indicated by the
              previous EFT computation. It also captures a similar constant $\eta$, which, as in the previous case, depends on the number of states. 
              \newline 
 As an example in the Enriques  CY $K3\times T^2/{ \mathbb{Z}_2}$, with $s$ the K\"ahler modulus of $T^2$
              one has  $c_2=12$ and 
              \beq
              V_{1-\rm{loop}}   \ \simeq  \  \frac {g^2\,m_{3/2}^2 M_{\rm P}^2}{(8\pi)^2} \left( 1\ -\  \frac {2\tilde{N_0}}{N} \ +....\right) \ .
              \label{potenri}
              \eeq
               This result has the same structure as the computations in section (\ref{section3}). 
                In fact we know that ${\tilde N}_0$ should include the number of massless chiral multiplets 
                denoted $N_0$ in section (\ref{section3}), and each chiral multiplet contributes 4 states
                to ${\tilde N}_0$, so that this dependence matches the $N_0$ dependence in eq.(\ref{V1loop}).
                Note however that 
               the computation  in \cite{vandeHeisteeg:2022btw}
               is sensitive to global properties of the CY compactification and the subleading terms 
               carry more information than just that of a particular one-modulus case as considered in section \ref{section3}.

                \subsection { Modular invariance in the one-loop potential}\label{oneloopmodular}
              
              The above computations tell us that at large modulus, the potential asymptotes to a cut-off independent 
              constant term \footnote{We will see momentarily that in fact in specific GKP vacua the gravitino mass is actually also moduli 
              dependent, asymptotically decreasing like some power of $1/s$. This will add additional modulus dependence to the 
              potential but will not change the location of the minima at the \textit{desert points}, see below.} $\sim m_{3/2}^2M_{\rm P}^2$. 
              Then the term $\sim -1/N$ will tend to drive the modulus to small values $s\sim 1$. However as we approach that limit, 
              one expects to have important corrections and one loses full control.  Still in some cases the dualities of the vacuum allow us
              to get information on the behaviour of the potential close to the self-dual points, as we now discuss.

               In  CY compactifications with a 2-torus factor, like in the Enriques CY or toroidal 
               orientifolds, the form of the species scale depending on vector moduli 
                may be obtained from the Topological String Theory computation in \cite{Bershadsky:1993ta}.
                              One has  \footnote{This functional form has also been shown to appear 
when computing threshold corrections in Type IIA  ${\cal N}=1$ toroidal $\mathbb{Z}_2\times \mathbb{Z}_2$ orientifolds,  see \cite{Blumenhagen:2007ip}.}
                 \begin{equation}
    N = -6 \log(2u |\eta(U)|^4 ) + \tilde{N_0}\ ,
    \label{N-toro}
\end{equation}
with $u={\rm Im}U$ the torus modulus.  
This expression is invariant under  SL$(2,\mathbb{Z})$ transformations on the torus field $U$.
   From here, one would then compute the one-loop potential from eq.(\ref{poten2}) and obtain 
   \beq
   \delta V \ \simeq \ \frac {g^2\,m_{3/2}^2M_{\rm P}^2}{(8\pi)^2}  \left(\frac {3a}{\pi}\right)^2 \left| \frac {u}{N} {\tilde G}_2(U,{\overline U} )\right|^2 \ ,
   \eeq
               Here      ${\tilde G}_2$ is the  Eisenstein (non-holomorphic)   weight-2  modular form. It may be written in terms of the holomorphic 
         (although non-modular covariant) weight-two as
         \beq
         {\tilde G}_2(U) = G_2  - \frac {\pi }{u} \  ;\  G_2 \ =\ \frac { \pi^2}{3} (1 -24e^{2\pi iU}\ +\dots ) \ .
         \eeq
One can easily show that any modular invariant function like the one above has extrema at the self-dual points $ U=\{i, e^{i\pi/3}\}$. 
In the present case, there are minima at both points with $\delta V_{min}=0$.  At the other extreme, in the large \( u \) limit, one finds the asymptotic expansions:
\begin{align}
N &\longrightarrow 2\pi u - 6\log(2u) + \dots \ , \\
G_2 &\longrightarrow \frac{\pi^2}{3} + \dots \ , \\
\Lambda &\longrightarrow (2\pi u)^{-1/2} \,.
\end{align}
Note that indeed $N$ is linear in $u$ for large modulus.
Substituting into the expression for the potential, one obtains (again, up to subleading corrections):
\begin{equation}
\delta V \simeq \frac {g^2\,m_{3/2}^2 M_{\rm P}^2 }{(8\pi)^2}\, \frac{a^2}{4} \left( 1 - \frac{2\tilde{N_0}}{N} + \dots \right),
\end{equation}
which, as expected, has the same form as the result in ~\eqref{potenri}.
We thus see that, in spite of $u\sim {\cal O}(1)$ being in a
region in which corrections are large, 
 modular invariance has allowed us 
to find out that there is a minimum of vanishing energy at a self-dual point. 

Let us note in passing that modular invariant EFTs in  4d ${\cal N}=1$ string vacua have a long history see
e.g.\cite{Ferrara:1989bc,Font:1990gx,Cvetic:1991qm,Lynker:2019joa,Leedom:2022zdm} and more recently in the context of the Swampland program  e.g.
\cite{Gonzalo:2018guu,Cribiori:2023sch,Casas:2024jbw,Chen:2025rkb}.

\section{Moduli fixing in \texorpdfstring{${\cal N}=1$}{Lg} 4d Type II orientifold models}\label{section6}

The fact that potentials for no-scale moduli are generated at one-loop generically (once SUSY is broken) may have important implications
for the problem of moduli fixing.  The standard procedure up to now in Type IIB  is to start with some GKP vacuum \cite{Giddings:2001yu} in which a flux potential fixes 
all complex structure moduli and the complex dilaton, and also a sufficient multiplicity of fluxes allows for a small 
superpotential $W_0$, so that one can eventually create a hierarchy. Then non-perturbative effects generate a superpotential which fixes an overall K\"ahler modulus $\rho$, see \cite{vonGersdorff:2005bf,Berg:2005yu,Westphal:2006tn,Cicoli:2008va,Cicoli:2008gp,Gallego:2017dvd,Antoniadis:2018hqy,AbdusSalam:2020ywo} for proposals in this regard. If this mechanism allows for a supersymmetric AdS vacuum, on top of this, one can further add
some anti-$D3$ branes in KS throats, allowing the theory to be uplifted to dS. This is the approach followed in the KKLT and LVS  \cite{Kachru:2003aw,Balasubramanian:2005zx,Conlon:2005ki}. In Type IIA,
more fluxes are available and one can fix all moduli in AdS using just the flux potential, as in  DGKT and CFI \cite{DeWolfe:2005uu,Camara:2005dc}. In that case, however is not clear how to 
perform an uplifting from AdS.

 Taking into account the above described one-loop potential contributions allows, in principle, for simpler new scenarios for moduli fixing.
 Let us consider for convenience the case of  Type IIA  ${\cal N}=1$ 4d  orientifold models,
 since we are using this class of models for the examples below. The same story would apply in Type IIB 
 orientifolds, doing an appropriate mirror transformation.  We start with a no-scale Minkowski vacuum in which the complex dilaton $S$ 
 and the K\"ahler moduli $T_a$ are fixed by a flux superpotential in the standard way. Classically the no-scale c.s. fields $U_i$, which do not
 appear in the superpotential, remain massless. They contribute to SUSY breaking as 
 $g_{i{\bar i}}
(D^iW)({\overline D}^{ {\bar i}}{\overline W}) \ = \ 3m_{3/2}^2$. A one-loop potential is generated for them in the way described in the
previous sections, so that minima occur at $u_i\sim {\cal O}(1)$. Thus this is potentially a very economical way to minimize all moduli 
at Minkowski.  In the Type IIB language, this would mean that there is no need for non-perturbative superpotentials to fix the K\"ahler (IIB) moduli
nor any uplifting
\footnote{ Alternatively,  the one-loop potential could play a role in the up-lifting process. For large $t$ moduli the one-loop potential behaves like a runaway potential of the form
$\delta V\sim |W_0|^2/t^r$, which is also the behaviour of the potential generated  by a anti-D-brane in  KKLT  and LVS.
In this case an additional minimum beyond the \textit{desert points} would be present for large moduli.
An exploration of the viability of this possibility would be worthwhile.}.

This is of course in principle a very interesting scenario. Still, like  in the classical  KKLT, LVS scenarios, there are some issues to be addressed,
some of which we enumerate:

{\it i)}
 The first is in fact shared by the KKLT or LV scenarios. It is non-trivial in general to fix {\it all } K\"ahler moduli (in IIA) and dilaton 
with fluxes and addressing RR-tadpole cancellation (in non-trivial theories) at the same time. This may be considered as 
a reflection of the ``RR tadpole'' conjecture of \cite{Bena:2020xrh,Bena:2021wyr}.  Also, a small $W_0$ should be obtained
by flux fine-tuning. 
In spite of these  difficulties, there are examples in which this is achieved, see e.g.\cite{McAllister:2024lnt} and
references therein.

{\it ii)} Perturbative control of the original classical vacuum requires the value of $s,t_a$  at the flux minima to be sufficiently large.
This issue also appears in scenarios like KKLT, LVS, DGKT and CFI. It may be in principle achieved by an appropriate choice of the fluxes. This is also constrained by RR-tadpole cancellations.

{\it iii)} Concerning the c.s. potential (in IIA), the one-loop potentials discussed above will fix those moduli at \textit{desert points}
with $u_i\sim {\cal O}(1)$. Thus perturbative control is lost in the c.s. sector, and non-perturbative corrections
could be large. The idea here is that the most relevant such corrections could already be included in the 
one-loop potential discussed above, which includes non-trivial dynamics in the bulk, particularly in the presence of duality 
symmetries like SL$(2,{\mathbb Z})$.

In this connection, it is interesting to compare the size  of the one-loop correction at hand  to 
the $\alpha'$ and one-loop corrections computed in the literature for the IIB case (thus for the K\"ahler moduli) at large modulus.
A useful systematic listing for perturbative corrections is provided in e.g. section (2.2) of ref.\cite{Cicoli:2021rub} (see also the analysis performed in \cite{Gao:2022uop}).  The terms 
are classified in terms of inverse powers of the volume ${\cal V}$ (recall that in IIB ${\cal V}^2 \sim T^3$).  The  $(\alpha')^3$ correction coming from  \cite{Becker:2002nn}
leads to a potential going  like  $W_0^2/{\cal V}^3$ and is the leading  $\alpha'$ correction. String loop corrections in Type IIB orientifolds in the presence of
D-brane moduli were computed in 2005 in  \cite{Berg:2005ja,Berg:2007wt}, see also \cite{vonGersdorff:2005bf}. These corrections scale like 
 $W_0^2/{\cal V}^{8/3}$ at large volume, and break the tree-level no-scale structure of the theory. In  \cite{Cicoli:2007xp} it was shown that there is
 an extended no-scale structure surviving such that these one-loop corrections become subleading.  
  On the other hand, additional ${\cal N}=1$ one-loop string corrections were computed for the same class of orientifolds in 2014 and in 2018 in Refs.~\cite{Berg:2014ama,Haack:2018ufg}. It was found there that corrections to the metrics of the Kähler moduli appear at one loop, thereby breaking the classical no-scale structure. Interestingly, these one-loop corrections to the metric are consistent with our computations in Section~\ref{section4}, where we showed that, in general, integrating out towers of states gives rise to such corrections to the moduli-space metric. More concretely, the one-loop correction under consideration, in the large-volume limit, yields (see Eq.~\eqref{metrica})
\begin{equation}
    \delta g_{i\bar{i}} \simeq e^{2\phi_4}\,\frac{1}{u_i^2} \,,
    \label{metric}
\end{equation}
(since $N\rightarrow u_i$) in perfect agreement with Eq.~(72) of Ref.~\cite{Haack:2018ufg}. Such a corrected metric lies at the origin of the large-moduli correction to the potential that we found, $\delta V \sim g^2\,m_{3/2}^2 M_{\rm P}^2$. At large volume, the one-loop correction to the potential computed in the present work scales as $W_0^2 / {\cal V}^3$, appearing as a subleading contribution compared to the tree-level term and of the same order as the $(\alpha')^3$ corrections computed in Ref.~\cite{Becker:2002nn}. Although these corrections are too suppressed to significantly affect the KKLT or LVS scenarios, the interesting point lies in what such corrections may imply for the small-moduli regime, which remains largely unexplored.

In the context of ${\cal N}=1$ supergravity it is interesting to see what type of correction to the K\"ahler
potential of the moduli of the theory could lead to
the correction we find. Starting from a metric is difficult (and ambiguous) to obtain a general Kähler potential
leading to that metric (as in eq.(\ref{metrica})). However one can be more specific along some moduli directions.
Thus e.g. at fixed 4d dilaton (which also means fixed string scale $M_s$) a Kähler potential giving eq.(\ref{metrica})
is
\beq
\delta K_{1-\rm{loop}}(g_4=cte) \  \simeq \  - \epsilon  \ \log N(T,{\overline T})  \ ,
\label{4ddil}
\eeq
Here  $\epsilon\sim 1/(8\pi )^2$ and $N \sim \Lambda^{-2}$, with $\Lambda$ the species scale moduli dependent function (with frozen  c.s.).
Since it is known that at large modulus $N\sim {\rm Im}T$, this correction is proportional in that limit to the tree level K\"ahler potential for $T$ and hence
destroys the no-scale tree level structure whose  `$-3$ factor'   gets modified  to $-(3+\epsilon)$. The correction is then large
since there is no field-dependent dilaton supression. 
However the field direction which is relevant in KKLT/LVS keeps the 10d IIB dilaton and complex structure fixed and the volume is large.
In that case the volume dependence pf the 4d dilaton becomes important and the correction is much smaller (for large 
Kähler moduli). In Ref.~\cite{Haack:2018ufg} the Kähler potential for large volume in a ${\mathbb Z}_6$ IIB orientifold coming from the 
corrected metric above was estimated to behave (to leading order) as
\beq
\delta K_{1-\rm{loop}}(g_{10}=\rm{const}) \sim \epsilon \ e^{2\phi_4} \ \sim \ \frac {\epsilon}{\sqrt{st_1t_2t_3}}
\label{10ddil}
\eeq
with $t_i$ the imaginary part of the Kähler moduli of the tori. As we said, this leads to quite small corrections to the KKLT/LVS 
models since the 'extended no-scale' property  is numerically respected. Note that at large volume the structures (\ref{4ddil}) and (\ref{10ddil}) look like
extreme limits  with behavior  $\sim  \log({\cal V})$ and $\sim 1/{\cal V}$. It would be interesting to investigate whether 
one can define a more global Kähler potential, with information also in the bulk.

The $\alpha'$ and loop corrections estimated so far in the literature
have only been computed for large modulus (large K\"ahler modulus in IIB), but nothing is known 
about the possible behaviour of such type of corrections as one approaches the \textit{desert points}. Two possibilities come to mind: i)  It could be that they vanish at those points, as it happens 
with the one-loop correction discussed in this paper. In this case the moduli will be fixed at the \textit{desert points} in a Minkowski vacuum; ii) It could be that they do not vanish, but they move 
the vacuum to dS or AdS.  Even in this case, note that the one-loop potential discussed here also appears even in AdS or dS vacua, see eq.(\ref{conV}).
Thus, the modulus will be stabilised, although we will not be certain whether the space remains in  Minkowski or not.

One may  argue that c.s. vevs of order one conflicts with phenomenology, since SM gauge couplings are functions of  $S$ and $U_i$
in these (IIA)  orientifolds, and they look perturbative at low energies.  However
$s$  may be large depending on the fluxes. On the other hand $u_i\simeq {\cal O} (1) $ is a boundary condition 
{\it at the string sale}. Renormalisation group running may make gauge couplings much smaller in the IR 
if the theory is not asymptotically free. This is in fact the case in many semi-realistic MSSM models, see Appendix \ref{apB}.

So in summary, to fix all moduli it would be enough to look for CY orientifolds fixing,  as usual,   $S$  and all  $T_a $ 
(in IIA case) via fluxes (respecting RR tadpoles), in a non-SUSY Minkowski vacuum, and with a small $W_0$. 
The one-loop potential will fix all vevs of no-scale c.s.  fields not present in the superpotential and contributing to SUSY breaking. A vacuum in Minkowski, which appears if the subleading corrections to the no-scale moduli vanish at the
\textit{desert points}, 
would be a good starting point to do phenomenology.  In fact the observed value of the c.c. is suspiciously close to the mass of neutrinos,  so very likely
one cannot ignore the SM sector of the theory while trying to obtain a theory with a c.c. close to observations.
Finally, the vacuum should not allow for instanton generated superpotentials for the c.c. which could destabilize the 
Minkowski structure.

  \subsection{ 4D  \mathtitles{{\cal N}=1} orientifold examples and moduli fixing in Minkowski }

      In the previous sections, we analysed the contribution of towers of states to the one-loop correction of the vacuum energy. We now turn to applying these results to some
      specific models for illustrative purposes. In the top-down examples here, we will discuss a few key differences compared to the previous sections.

First, as we have already advanced, we will see that the UV 'constant' contribution identified earlier becomes moduli-dependent (an outcome that one would naturally anticipate in a theory of quantum gravity). 
This is because in a no-scale theory, the gravitino mass is field-dependent but also the 4d dilaton is.
Second, we will recover the usual result that for large values of the fields, the one-loop effective potential asymptotically approaches zero. This behaviour is in line with expectations from the Swampland Distance Conjecture \cite{Vafa:2005ui} and the de Sitter Conjecture \cite{Obied:2018sgi,Ooguri:2018wrx}, which suggest that extended flat plateaux in moduli space are difficult to achieve in consistent theories of quantum gravity.
\newline
Finally, we will find that away from the infinite-distance regime, the generated one-loop potential can lead to moduli stabilization both in the bulk of moduli space and at a dS maximum, illustrating a mechanism of potential moduli fixing that is compatible with these Swampland ideas.

      We will use as our laboratory the well-known toroidal orientifold $\mathbb{Z}_2\times \mathbb{Z}_2$ in 4d ${\cal N}=1$, described in detail in e.g. \cite{Camara:2005dc}.  It is obtained by compactifying Type IIA on  a compact toroidal 
      space ${\bf X}_6 = ({ T}^2)_1\times  ({ T}^2)_ 2 \times  ({ T}^2)_3 / {\bf \Gamma}$ where ${\bf \Gamma}$ is a  $\mathbb{Z}_2\times \mathbb{Z}_2$
      orbifold action, leading to a 4d ${\cal N}=2$  theory. To be more precise, we will choose the version of this orbifold with discrete torsion \cite{Font:1988mk}, as in \cite{Marchesano:2004xz},
      which has $h_{11}=3$, $h_{21}=3+48=51$.
       To obtain a ${\cal N}=1$ theory one makes an orientifold projection 
      by $\Omega_p(-1)^{F_L}\sigma$, where $\Omega_p$ is the world-sheet parity operator, $F_L$ is the space-time fermionic number for
      left-movers and $\sigma $ is an order 2 reflection operator leaving fixed $O(6)$ orientifold planes. In particular for tori coordinates $z_i=x_i+i\tau_iy_i$,
     the operator $\sigma$ will act as $\sigma(x_i)=x_i$, $\sigma(y_i)=-y_i$. Here $\tau_i=R_y^i / R_x^i$  is the {\it geometric}  complex structure. The untwisted complex structure
      fields have imaginary parts (saxions) given by
      \beq
      s = \frac {e^{-\phi_4}} {\sqrt{\tau_1 \tau_2 \tau_3}} \ , \quad
      u_i =  e^{-\phi_4} \sqrt{\frac {\tau_j \tau_k}{\tau_i}} \ ,\ i\not=j\not= k \ .
      \eeq
      The 4d dilaton is defined  by $e^{-2\phi_4}=4\sqrt {su_1u_2u_3}$, and one has for the string scale $M_s=e^{\phi_4}M_{\rm P}$.
      One also has
      \beq
      \tau_i \ =\ \sqrt{\frac {u_ju_k}{su_i}   }\ .
      \eeq
      The saxions of the untwisted K\"ahler moduli parametrise the area of the tori, $t_i={\rm Im}T_i = A_i=R_y^iR_x^i/4$. For $\tau_i\ll 1$ there are KK and winding
      towers  with scales
      \beq
      m_{\rm{KK},i}^2 \ =\   \frac {M_{\rm P}^2} {4st_iu_i} \ ,\quad  m_{\rm{w},i}^2 \ =\   M_{\rm P}^2  \frac {t_i}{4su_i} \ .
      \eeq
      Also, tensionless strings may appear is some limits from  D$4$-branes wrapping 3-cycles or  NS5 wrapping 4-cycles, with tensions
      \beq
      T_{\rm{D}4}^i \ = \ \frac {M_{\rm P}^2}{4u_i} \  , \  T_{\rm{NS}5}^i \ = \ \frac {2M_{\rm P}^2}{t_i} \ .
      \eeq
       The leading order K\"ahler potential for untwisted moduli is given by the well-known 
       leading order expression
       \beq
       {\cal K}\ =\ -\log(-i(S-S^*))\ -\ \sum_i \log(-i(U_i-U_i^*)) \ -\ \sum_i \log(-i(T_i-T_i^*)) .
       \label{Kahler}
       \eeq
        Adding NS and RR fluxes, superpotentials are generated. Allowing for metric fluxes the general expression of the flux
        superpotential is    (see e.g., \cite{Camara:2005dc} and references therein):
        \beqa
        W_{\rm flux}   \ &=& \ e_0 \ + \ h_0S \ +\ \sum _{i=1}^3[T_i( e_i \ +\ a_iS \ -\ b_{ii}U_i\ + \ \sum_{j\not=i}b_{ij}U_j)     \ + \ h_iU_i] \notag\\
        &+& \  q_1T_2T_3 \ +\ q_2T_1T_3 \ +\ q_3T_1T_2 \ -\ mT_1T_2T_3 \ .
        \eeqa
          Here $e_0,e_i,q_i,m$ are respectively $F_6,F_4,F_2$ and $F_0$ RR fluxes and  $h_0,h_i$ are $H_0^{(3)},H_i^{(3)}$ NS fluxes.  
          In addition, $a_i,b_{ij}$ are metric fluxes
          see e.g.\cite{Derendinger:2004jn,Villadoro:2005cu,Camara:2005dc}.
          All of them are integrally quantized, as we discuss later.
          We will not consider here the issue of twisted moduli fixing for simplicity. Fixing of twisted moduli 
        may be obtained e.g., by adding twisted fluxes as in \cite{Narayan:2010em}  or, in the presence of D-branes, combining soft induced mass terms for charged fields with twisted Fayet-Iliopulos \cite{Cascales:2003wn,Blumenhagen:2005tn,GarciadelMoral:2005js}.
          
          We will be interested first in constructing a no-scale model in which the vevs of  $u_i$ remain undetermined.
          Thus for simplicity we will take $e_0=e_i=h_i=b_{ij}=0$ so that we are left with the superpotential
          \footnote{Note that in the Type IIB mirror the $a_i$ fluxes map into regular $H^{(3)}$ 3-form fluxes.}
          \beq
        W_{\rm flux}   \ =  \ h_0S \ +\ \sum _{i=1}^3 a_iST_i   \ +\   q_1T_2T_3 \ +\ q_2T_1T_3 \ +\ q_3T_1T_2 \ -\ mT_1T_2T_3 \ .
        \eeq
        This has no dependence on the 3 $U_i$ moduli, as required to get a no-scale model. A number of no-scale examples may be constructed, as described, e.g., in \cite{Camara:2005dc}. 
        Let us briefly discuss two of these here.

        {\it i) A model with all moduli fixed in Minkowski.}
        One can see that with this superpotential, a no-scale solution is obtained
        \cite{Derendinger:2004jn,Camara:2005dc}, e.g., in the isotropic case with $a_i=a$, $q_i=q$, with Minkowski minima at
        $t_i=t$ with 
        \beq
         t\ = \ \sqrt{\frac {h_0q}{ma}} \ ,\ s\ = \ \frac {q}{a} t \ .
         \eeq
        and
        \beq
        m_{3/2}^2 \ =\ \frac {(h_0m+9qa)}{32u_1u_2u_3} \ =\   (h_0m+9qa) \frac {M_s^4}{2M_{\rm P}^2} \ .
        \eeq
        This corresponds to the background labelled NS4 in ref.\cite{Camara:2005dc}.  On the other hand, the c.s. fields will have locally a one-loop 
        potential (with $t,s$ fixed at their vevs)
        \beq
        V_u\ \simeq \  \frac {2\,e^{2\phi_4}\,M_s^4}{(8\pi)^2}  (h_0m+9qa) \frac {1}{3} \sum_{i=1}^3 \frac {\left|\partial_{U_i} \Lambda (U_i,{\bar U}_{\bar i}) \right|^2}{\Lambda^2} \ ,
        \eeq
        where $\Lambda$ is the species scale at fixed $S,T_a$.
       For large complex structure the potential is runaway. Thus along the $u_i\rightarrow \infty $ limit, leaving the other two $u_ju_k$, $i\not=j\not=k\not=i$
       at their minima of order one, one has a one-loop potential 
       \beq 
       V(u_i)\ \longrightarrow \  \frac {(h_0m+9qa)}{48\,(8\pi)^2}  \frac {1}{u_i^{3/2}} \ .
       \eeq
        This is a runaway potential, which is consistent with Swampland expectations.  On the other hand, as discussed in previous sections, 
        all three moduli will have minima at the \textit{desert points} at which $\partial_i\Lambda=0$,   and vanishing potentials.
        \footnote{In these vacua the 4d dilaton behaves like $e^{4\phi_4}\sim 1/(16 s)$. Thus one can get a small 4d dilaton by
        chosing fluxes so that $s$ is large. RR tadpole conditions may however restrict how large $s$ can be made.}
        This would be a neat and simple example of a model with all moduli fixed quite economically. Unfortunately, as we discuss further below,
        it is difficult to cancel all RR tadpoles in this particular model, since all flux quanta in this orientifold must be a multiple of 8, see refs \cite{Blumenhagen:2003vr,Cascales:2003zp,Cascales:2003pt} Still, it shows us a way in which 
        full moduli fixing could take place in a model less restricted by tadpoles.

        {\it ii) A model with a modular invariant potential}
        
        As we discuss above, one has explicit expressions for the species scale in ${\cal N}=2$ theories with a CY compactification involving a torus $T^2$, like
        the Enriques surface $K3\times T^2/\mathbb{Z}_2$ or even the present $\mathbb{Z}_2\times \mathbb{Z}_2$ example. However, only the piece involving 
        vector multiplets is under full control and has the neat expression in \eqref{N-toro}. When orientifolding one expects that one has a 
        similar expression for the ${\cal N}=1$ case concerning K\"ahler moduli in IIA or c.s. in IIB. Unfortunately, those are the moduli which may be 
        directly fixed by fluxes anyway and furthermore have vanishing auxiliary field so that they do not get any one-loop potential of the type described above. 
        Thus we are unable to write an explicit SL$(2,\mathbb{Z})$ invariant potential for the complex structure fields
        (in IIA) without further input about the hypermultiplet dependence of the species scale \footnote{    
        In some simple cases the generic instanton effects affecting the hypermultiplet dependence   may be absent or suppressed. Thus in e.g. the IIA ${\mathbb{ Z}}_2\times \mathbb{ Z}_2$ orbifold {\it without}  discrete torsion
one has  that the only  three hypermultiplet   fields come  from the untwisted sector. That sector is known to be uncorrected by  instanton effects.
 In this case to leading order the species scale for
c.s  moduli  would be the mirror of the K\"ahler  one, see e.g.\cite{Gregori:1997hi} and references therein.
 For  generality we will keep in what follows the species scale for the  IIA  c.s. moduli unspecified.}.

         One may write however an example of a no-scale model in which one of the K\"ahler moduli, say $T_1$ is a no-scale field and does not appear in the 
         flux superpotential.  Consider the flux superpotential
         \footnote{Note that this is not then a GKP class of model since both K\"ahler and c.s. moduli appear in the superpotential.}
         \beq
        W\ =\ S(a_2T_2+a_3T_3) \ +\ U_1(b_{21}T_2 + b_{31}T_3) \ .
        \eeq
       Here $a_i$ and $b_{ij}$ are metric fluxes.  The no-scale moduli are now $T_1,U_2,U_3$ \footnote{This is a variant of the NS2 model of ref.\cite{Camara:2005dc} with the 
       replacements $U_1\leftrightarrow T_1$, $q_3\leftrightarrow b_{21}$ and $q_2\leftrightarrow b_{31}$.}.  One can check that there is a Minkowski 
       no-scale minimum with all axions in $S,U_1,T_2,T_3$ vanishing and saxions constrained by
       \beq
       s^2 \ =\ \frac {q_2b_{21}} {a_2a_3} u_1^2 \ ;\ t_3^2 \ =\ \frac {a_2b_{21}} {a_3b_{31}}t_2^2 \ .
       \eeq
        Thus in this case there is no full  $s,u_1,t_2,t_3$ moduli fixing but only two constraints. 
        On the other hand, we  have as one of the three  no-scale fields the K\"ahler modulus $T_1$ for which one can write the
        corresponding one-loop potential  in a quite explicit manner in terms of an SL$(2,\mathbb{Z})$ modular invariant species scale 
        as described in section 
        \ref{oneloopmodular} for the case of the $T^2$ torus. Thus,  setting $u_2,u_3$ at their 
        desert point minima and moving along the direction $\phi_4$ is fixed,   the $T_1$-dependent potential piece  has the form
        \beq
        V_{t_i,u_s,u_3}\ \simeq \  4\, \frac {(a_2b_{31}+a_3b_{21})}{(8\pi)^248(t_1)}  
         \left(\frac {6}{\pi}\right)^6\left|\frac {t_1}{N}{\tilde G}_2(T_1,{\bar T}_{\bar 1})\right|^2\,M_{\rm P}^4  \ .\label{Vt1}
                \eeq
 Note that this expression, as it stands, applies down to $t_1\geq 1$, i.e., one restricts oneself to the SL$(2,\mathbb{Z})$ fundamental domain of $t_1$. Recall that the gravitino mass is computed in the perturbative regime and does not display modular invariance 
 due to the gravitino mass dependence on  $t_1$. One would naturally expect equation \eqref{Vt1} to be completed in a modular
 invariant way with respect to $t_1$ close to the \textit{desert point}. Still one does not expect such a completion to modify
 the structure and the existence of the minimum, since the gravitino mass should not diverge at the desert point.

 As we said in this particular $\mathbb{Z}_2\times \mathbb{Z}_2$ orientifold model, the fluxes are quantized in multiples of 8, making it difficult to cancel tadpoles.  It turns out however that the situation improves a lot if we add D$6$-branes, as we describe in appendix \ref{apB}.
We discuss there, some MSSM-like examples with RR tadpole cancellation, and in which the c.s. moduli $U_i$ are fixed at the 
\textit{desert points} and partial K\"ahler/complex dilaton moduli fixing.

Another important point to emphasise is that in these toroidal models, which only have a few types of fluxes, the gravitino mass is close to the string scale. 
In order to get a hierarchy, one would need a model with a rich flux structure so that one
can tune a small $W_0$.

\section{ Comments on UR-IV correlations and inflation}\label{section7}

In the computation of these one-loop potentials two interesting points arise which are worth remarking.

{\it IR-UV correlations.}
The first of them is that these one-loop potentials represent an explicit example of UV-IR  mixing effects in the following sense.
Looking from the IR point of view one just sees the light modes and a  quadratically divergent negative contribution to 
the potential. One would thus say naively that the theory predicts a large negative contribution 
$V\sim -m_{3/2}^2M_{\rm P}^2$ to the cosmological constant.
 However this is not correct for two reasons. First,  the modulus comes along 
with a large tower of massive states, each of them giving a positive contribution to the vacuum energy, and second,
because the UV cut-off is not the Planck scale but rather the (field dependent) species scale. 
When putting light (IR) and tower (UV) states together,  they conspire to yield a cancelling vacuum energy 
at the minimum of the full one-loop potential.  From the point of view of the IR states this is miraculous and unexpected,
but it makes perfect sense when considering the full pack of light and tower states. The cut-off is the species scale $\Lambda$, which in this case is
very close to the Planck scale, since the minimum is at a desert point. The cosmological constant vanishes
at the minimum, but the gravitino mass is independent of this fact, and its value will depend on the value
of $W_0$, i.e., the flux configuration.

The way this UV-IR correlation happens is very intriguing and suggests that perhaps some similar effect may happen when computing loop corrections at the EW scale. 
From the IR point of view in the SM, the Higgs particle gets a large quadratically divergent contribution to its mass and also a huge contribution
to the cosmological constant. One could conceive that a similar mechanism as above could be relevant, with the c.c. remaining small even after EW symmetry breaking. 
It would be interesting to study whether one can implement this idea more  concretely

{\it Inflation}

In ref.\cite{Casas:2024jbw} we proposed a SL$(2,\mathbb{Z})$ modular invariant potential as a candidate for inflation. Our guidelines for the proposal 
were to have an asymptotic plateau in order to get sufficient inflation, modular invariance on the (complex) inflaton,
and dependence just on the species scale and their derivatives. It had the form
\beq
V \ =\ \mu^2M_{\rm P}^2 g^{\tau{\bar \tau}}\frac {(\partial_\tau \Lambda)(\partial_{\bar \tau}\Lambda)}{\Lambda^2} \ ,
\eeq
where $\mu$ is a constant mass scale to be adjusted to the amplitude of perturbations. 
We did not give any justification for the origin of such a potential from string theory. 
Now we see that, interestingly,  it corresponds to the one-loop induced potential of a no-scale modulus if
we identify $\mu^2=e^{2\phi_4}\,m_{3/2}^2$.
We used as a proxy for the species scale the toroidal one in eq.(\ref{N-toro}). Then expanding on $1/\tau_2$ 
and going to a canonical frame for $\tau_2$ one gets (up to log corrections) \cite{Casas:2024jbw}
\beq
V\ \longrightarrow \mu^2 M_{\rm P}^2 \frac {9}{2\pi^2}\left|\frac {\pi}{3} \ -\ e^{-\sqrt{2/3}\,\phi }\right|^2 \ .
\eeq
Interestingly, for $\mu$ constant, this has the structure of a Starobinsky inflaton potential, which is one of the
most successful inflaton potentials used in the community. These types of potentials have been further generalized later on 
by Kallosh and Linde in \cite{Kallosh:2024ymt,Kallosh:2024kgt,Kallosh:2024pat} and also in \cite{Carrasco:2025rud}. 
On the other hand, in the context of the present paper, 
the gravitino mass depends in general on the no-scale saxions like $m_{3/2}^2\ \sim \ 1/u^r$,  and $e^{2\phi_4}\sim  1/\sqrt{su^r}$ for a positive $r\leq 3$.
Then the potential has the general form of the orange curve with a bump in Fig.\ref{fig:potentials} rather than the blue one with a plateau. Still, it should be interesting to check whether even with a field-dependent gravitino mass 
one can obtain sufficient inflation going downhill towards the \textit{desert points}. We leave this work for the near future.

\section{Discussion}\label{conclus}

In this paper we have pointed out that no-scale moduli in a spontaneously broken  ${\cal N}=1$, 4d Type II vacua, which are massless at tree level, 
get a positive definite potential at one-loop. The potential vanishes at the {\it desert points } and at infinity, with a dS plateau in 
between with a height of order ${\cal O}(g^2\,m_{3/2}^2M_{\rm P}^2)$, with $g$ the loop-counting coupling parameter $g^2=e^{2\phi_4}$. The one-loop potential is contributed to by both the IR particles
with mass $\lesssim m_{3/2}$ and the infinite tower of states which become massless at large modulus. In the computation, a
crucial point is that one uses the species scale as a cut-off, and also one takes into account the moduli dependence of this cut-off. 
An important cross-check of our results is that the structure of the one-loop corrections at large moduli is consistent with the findings of Ref.~\cite{Berg:2014ama,Haack:2018ufg} for the one-loop corrections to the Kähler metrics in four-dimensional $\mathcal{N}=1$ theories. In particular, the scaling behavior and functional dependence we obtain follow the same pattern as those derived in explicit string computations, providing further evidence for the robustness of our approach.

We argue that this one-loop potential has Minkowski minima at the {\it desert points} in moduli space, which correspond to the maxima of the 
species scale. This drives the no-scale moduli vev towards these \textit{desert points}, giving rise to what we call 'moduli self-fixing'.
This self-fixing applies only to the moduli contributing to SUSY-breaking (and hence to the gravitino mass).  Thus e.g., 
in the case of Type IIB  CY orientifolds gives rise to moduli fixing of K\"ahler moduli. This may have important implications
for the issue of moduli fixing in general. Thus, e.g., in a KKLT type setting, non-perturbative superpotentials would not be needed
to fix the no-scale K\"ahler moduli. 

Although we know there should be minima at the \textit{desert points}, it is difficult to get complete control in this
non-perturbative region of moduli space. The potential is determined by the gravitino mass and the gradient of $\log(\Lambda)$, thus, we would 
require full knowledge of the species scale to have access to the precise form of the potential close to the \textit{desert points}. We have argued that 
the dualities of string theory, which are encoded in the species scale, could help us to get this access. We have shown an example involving 
$T^2$ tori in the compactifications. It would be important to extend this idea to more general settings.

We have presented a possible path to achieve full moduli fixing in Type II CY orientifolds by combining RR-NS fluxes with the self-fixing one-loop potentials
and illustrated it  with  a $\mathbb{Z}_2\times \mathbb{Z}_2$ toroidal orientifold example.
Although only in some cases one has an explicit  (e.g., SL$(2,\mathbb{Z})$ invariant) form for the species scale, one can argue for full moduli fixing,
modulo the RR-tadpole condition, which is, as usual in any setting, hard to obtain. 

Most (almost all ) searches for moduli fixing up to now have tried to have full control over the perturbative parameters of the vacuum.
We all love to have perturbative control over the couplings. But Nature may think otherwise and prefer some (or all) of the
moduli to be at or close to ``self-dual'' (desert) points, as happens for the one-loop potentials described in this paper. Then we will have to get 
used to it and make use of new tools, like dualities. As explained above, having such a strong coupling regime in the UV is not in conflict 
with having perturbative SM couplings in the IR, if the MSSM model is extended so that it is non-asymptotically free, as often happens in specific constructions.

\vspace*{.8cm}

       \centerline{\bf \large Acknowledgements}

\vspace*{.5cm}
We would like to thank Michele Cicoli, Arthur Hebecker and Fernando Quevedo for their illuminating email exchanges about  the structure of one-loop string corrections in orientifolds. We also thank  Christian Aoufia, Alberto Castellano, Anamaria Font, Michael Haack, Fernando Marchesano, Miguel  Montero, Angel Uranga, and Max Wiesner for discussions. This work is supported through the grants CEX2020-001007-S and PID2021-123017NB-I00, funded by MCIN/AEI/10.13039 /501100011033 and by ERDF A way of making Europe. G.F.C. is supported by the grant PRE2021-097279 funded by MCIN/AEI/ 10.13039/501100011033 and by ESF+. 


\appendix

\section{One-loop effective potential}
\label{appendix one loop}

   The one-loop effective potential for a particle in
$d$ dimensions is given by \cite{Schwartz_2013}
\begin{equation}
V_{\mathrm{eff}}^{(d)}=(-)^{F+1} \frac{n_{\mathrm{dof}}}{2^{d+1} \pi^{d / 2}} \int_{s_0}^{\infty} \frac{d s}{s^{\frac{d}{2}+1}} e^{-s m^2}=(-)^{F+1} \frac{n_{\mathrm{dof}}}{2^{d+1} \pi^{d / 2}} s_0^{-d / 2} E_{\frac{d}{2}+1}\left(m^2 s_0\right), \label{eq: oneloopEd}
\end{equation}
where $F$ is the fermion number, $n_{dof}$ counts the degrees of freedom and $E_{\frac{d}{2}+1}$ is the exponential integral function. The expansion of $E_{\frac{d}{2}+1}$ for small $z$ is
\begin{equation}
E_n(z)=\frac{(-z)^{n-1}}{(n-1)!}(\psi(n)-\ln z)-\sum_{\substack{k=0 \\ k \neq n-1}}^{\infty} \frac{(-z)^k}{k!(1-n+k)},
\end{equation}
with $\psi(n)$ being the Digamma function. Substituting for $d=4$ and truncating to the first three terms, we arrive at 
\begin{equation}
    E_3(z) = \frac{z^2}{2}\left(\psi(3)-\log z\right) \, + \frac{1}{2} - z \, + \mathcal{O}(z^3).
\end{equation}
Therefore, putting altogether,  for e.g., the contribution of scalars in the loop, one gets
\begin{equation}
    V_{b} \simeq  \frac{1}{2^6 \pi^2}\left(-\Lambda^4 + 2 m^2\Lambda^2 -m^4\log\left(\frac{\Lambda^2}{m^2}\right) - m^4\psi(3) \right) + \mathcal{O}\left(\frac{m^6}{\Lambda^6}\right).
    \label{Vb}
\end{equation}
Since $\psi(3)=0.922784$, we can approximate the one-loop potential for bosonic particles as
\begin{equation}
    V_{b}^{\rm eff} \simeq  \frac{1}{2^6 \pi^2}\left(-\Lambda^4 + 2 m^2\Lambda^2 -m^4\log\left(\frac{\Lambda^2}{m^2}\right)  \right).
\end{equation}
We are interested in supersymmetric theories in which both bosons and fermions circulate in the loop. The total one-loop potential is then given by the sum over all species \cite{PhysRevD.7.2887,PhysRevD.7.1888}
\begin{equation}
    V_{\rm total}^{\rm eff} = \frac{1}{2^6 \pi^2}\left(-\text{Str}\mathcal{M}^0\Lambda^4 + 2\,{\rm Str}\mathcal{M}^2\Lambda^2 - 2\, {\rm Str}\mathcal{M}^4 \log\left(\frac{\Lambda}{\mathcal{M}}\right)\right)
\end{equation}
where we have defined
\begin{equation}
    {\rm Str}\mathcal{M}^a = \sum_n (-1)^{2j_n}(2j_n + 1)( m_n)^a,\quad a=0,2,4,
\end{equation}

\subsection{Str\mathtitles{\mathcal{M}^4} computation}\label{app: strm4}

In Section~\ref{section3}, we derived the one-loop effective potential that arises in supersymmetric theories. The general expression takes the form
\begin{equation}
    V_{\rm total}^{\rm eff} = \frac{1}{2^6 \pi^2}\left(-\text{Str}\mathcal{M}^0\Lambda^4 + 2\,{\rm Str}\mathcal{M}^2\Lambda^2 - 2\, {\rm Str}\mathcal{M}^4 \log\left(\frac{\Lambda}{\mathcal{M}}\right)\right), \label{oneloop}
\end{equation}
where the supertraces $\text{Str} \mathcal{M}^k$ are defined by
\begin{equation}
\text{Str} \mathcal{M}^a = \sum_n (-1)^{2j_n}(2j_n + 1)(m_n)^a.
\end{equation}
As discussed in the main text, the dominant contribution to the potential comes from the second term in Eq.~\eqref{oneloop}. The first term vanishes in theories with spontaneous supersymmetry breaking, and the third term, involving $\text{Str}  \mathcal{M}^4$, is typically subleading compared to $\text{Str}  \mathcal{M}^2$.
\newline
In the remainder of this appendix, we explicitly compute this contribution and demonstrate that, for towers of states with varying values of $p$, the term involving $\text{Str}  \mathcal{M}^4$ remains suppressed.

The general expression of this term for the chiral states in a tower of mass $m$, after SUSY breaking, is given by 
\begin{equation}
     {\rm Str}\mathcal{M}^4 \log\left(\frac{\Lambda}{n^{1/p} m}\right) = \frac{2}{p} \sum_n \left[(n^{2/p}m^2 + m_{3/2}^2)^2 - n^{4/p} m^4\right]\log\frac{N}{n}.\label{strm4ap}
\end{equation}
Let us compute this expression for different $p$'s:
\vspace{1em}

\subsubsection*{\underline{$p=1$}}
\begin{align}
    {\rm Str}\mathcal{M}^4 \log\left(\frac{\Lambda}{n m}\right) =& 2 \sum_n \left((n^2m^2 + m_{3/2}^2)^2 - n^4 m^4\right)\log\frac{N}{n}\\
    &\simeq 4 \sum_n n^2 m^2 m_{3/2}^2 \log\frac{N}{n} = 4 m_{3/2}^2 m^2 \sum_n n^2\log N - n^2\log n \notag \\
    & 4 m_{3/2}^2 m^2\left(\frac{1}{6}N(N+1)(2N+1)\log N + \zeta^{'}(-2)-\zeta^{(1,0)}(-2,N+1) \right)\notag
\end{align}
Expanding for large $N$, and taking into account that the characteristic mass (in Planck units) of the tower $m$ scales as $m^2\simeq N^{-3}$ with the number of species, we obtain,
\begin{equation}
   m_{3/2}^2\left( \frac{4}{9}-\frac{1}{3 N^2}+\mathcal{O}(N^{-4})\right).
\end{equation}

\subsubsection*{\underline{$p=2$}}

\begin{align}
    {\rm Str}\mathcal{M}^4 \log\left(\frac{\Lambda}{n^{1/2} m}\right) =&  \sum_n \left((n m^2 + \mu^2)^2 - n^2 m^4\right)\log\frac{N}{n}\\
    &\simeq 2 \sum_n n m^2 \mu^2 \log\frac{N}{n} = 2 \mu^2 m^2 \sum_n n \log N - n\log n \notag\\
    & 2 \mu^2 m^2\left(\frac{1}{2} N (N+1) \log (N)-\log (H(N)) \right) \notag
\end{align}
where $H(N)$ is the hyperfactorial. The hyperfactorial series expansion is given by,
\begin{equation}
   \left(
  A N^{1/12}
  + \frac{A}{720} N^{-23/12}
  + O\left(N^{-37/12}\right)
\right)
e^{\left(
  \frac{1}{4} \left(2 \log N - 1\right) N^2
  + \frac{N}{2} \log N 
  + O\left(N^{-4}\right)
\right)}
\end{equation}
where $A$ is the Glaisher constant. At leading order we obtain, 
\begin{equation}
     {\rm Str}\mathcal{M}^4 \log\left(\frac{\Lambda}{n^{1/2} m}\right)\sim 2 \mu^2 m^2\left(\frac{1}{4}N^2 + \mathcal{O}\log (N^{})\right)\simeq \frac{1}{2}\mu^2 M_{\rm P}^2
\end{equation}
{\underline{String limit}
\vspace{1em}

One could borrow the expression \eqref{strm4ap} and take the limit for infinite $p$, which yields
   \begin{equation}
     {\rm Str}\mathcal{M}^4 \log\left(\frac{\Lambda}{n^{1/p} m}\right) \simeq   lim_{p\rightarrow \infty}   \,\frac{2}{p} \sum_n \left[(n^{2/p}m^2 + m_{3/2}^2)^2 - n^{4/p} m^4\right] \log\left(\frac{\Lambda}{ \Lambda}\right) =0\ .
\end{equation}
However, the actual computation involves summing the string states undergoing the Hagerdorn degeneration. One starts with
\begin{equation}
    \sum_n^{N} d(n) \,\,n\,m_n^2 \,\, m_{3/2}^2\, \log\frac{\Lambda}{n^{1/2}m_n}
\end{equation}
where we can already substitute $m_n=M_s\,$, $d(n)\sim e^{\sqrt{n}}$ and $\Lambda/M_s = N_t^{1/2}$. Therefore, one arrives at
\begin{equation}
  \frac{1}{2}  M_s^2\,m_{3/2}^2\, \sum_n^{N} e^{\sqrt{n}}\,n\,\log\frac{N}{n}.
\end{equation}
The asymptotic result yields
\begin{equation}
    \sum_n^N e^{\sqrt{n}}\,n\,\log \frac{N}{n}\sim 4 Ne^{\sqrt{N}}
\end{equation}
Therefore, putting altogether
\begin{equation}
    2\,M_s^2\,m_{3/2}^2\,Ne^{\sqrt{N}}
\end{equation}
Using that \cite{Castellano:2022bvr},
\begin{equation}
   \left( \frac{M_{\rm P}}{M_s}\right)^2 \simeq 2 N^{3/2}\,e^{\sqrt{N}},
\end{equation}
we obtain
\begin{equation}
   M_{\rm P}^2\, m_{3/2}^2\,\frac{1}{N^{1/2}}\simeq  M_{\rm P}^2\, m_{3/2}^2\,\frac{1}{\log N_t},
\end{equation}
which is a log correction.

Therefore, based on the previous results, we can confirm that the important term in the one-loop potential is given by Str$\mathcal{M}^2$, up to subleading corrections from higher terms in the expansion of \eqref{eq: oneloopEd}

\section{Adding semi-realistic D6-branes to cancel RR tadpoles}\label{apB}

As we said in this particular $\mathbb{Z}_2\times \mathbb{Z}_2$ orientifold model, the fluxes are quantised in multiples of eight, which makes it difficult to cancel out tadpoles. One can slightly improve the situation by the addition of D6-branes. 
This orientifold  contains four types of $O(6)$ planes and the structure of the tadpole conditions for the choice of fluxes of the
previous section (see \cite{Camara:2005dc} for notation ) is
\beq
\sum_a N_a n_a^1n_a^2n_a^3\ +\ \frac {1}{2} (h_0m + a_1q_1+a_2q_2+a_3q_3 ) \ = \ 16
\eeq
\beq
\sum_a n_a^1m_a^2m_a^3=\sum_a m_a^1n_a^2m_a^3=\sum_a m_a^1m_a^2n_a^3= \ =\ -16 \ .
\eeq
Here $n_a,m_a$, $a=1,2,3$  are integer numbers giving the wrapping numbers of the D$6_a$-branes around the $x-y$ directions of the three
tori. For simplicity, it is assumed that all D$6$-branes pass through the sum $\mathbb{Z}_2\times \mathbb{Z}_2$ orbifold point. Each stack of N  D$6$-branes 
(and orbifold images) gives rise to a gauge group $U(N/2)$, so $N$ must be even \cite{Marchesano:2004xz}.

Let us consider the set of D6-branes with wrapping numbers as in table
(\ref{tablaSM}) \cite{Marchesano:2004xz}.   One can see that the massless spectrum of chiral multiplets 
at the intersections  (excluding singlets for simplicity of presentation) is that of   (\ref{espectro}). So it consists of three generations of 
a Pati-Salam group containing the SM fields and a number of $SU(2)_L,SU(2)_R$ doublets.  
 It is easy to check that, in the absence of fluxes, all RR-tadpoles cancel for $N_f=40$.  Note that the additional branes $h_1,h_2$ are 
 crucial because they precisely cancel the 3 twisted tadpoles, and their contribution to the untwisted one  $(n_a^1n_a^2n_a^3)$  is negative 
 and equal to. $(-96)$, allowing for a larger contribution from fluxes.

 \begin{table}[h!!]
 \begin{center}
			\renewcommand{\arraystretch}{1.00}
			\begin{tabular}{|c|c|c|c|}
				\hline
				$N_\alpha $  &  $(n_\alpha^1,m_\alpha^1)$  & $(n_\alpha^2,m_\alpha^2)$  & $( n_\alpha^3,m_\alpha^3)$  \\
				\hline\hline
				$N_a=6+2 $ &  $(1,0)$  &   $(3,1) $  &  $(3,-1) $\\ 
				\hline
				$ N_b=2 $ &  $(0,1)$  &   $(1,0) $  &  $(0,-1)$\\ 
				\hline
				$N_c=2 $ &  $(0,1)$  &   $(0,-1) $  &  $(1,0)$ \\ 
				\hline
				$N_{h_1}=2 $ &  $(-2,1)$  &   $(-3,1) $  &  $(-4,1)$ \\ 
				\hline
				$N_{h_2}=2 $ &  $(-2,1)$  &   $(-4,1) $  &  $(-3,1)$ \\ 
				\hline
				$N_{f}$   &  $(1,0)$  &   $(1,0) $  &  $(1,0)$\\ 
				\hline	
			\end{tabular}
			\caption{ D$6$-brane wrapping numbers of a Type IIA $\mathbb{Z}_2\times \mathbb{Z}_2$ orientifold 
			with MSSM-like spectrum.}
			\label{tablaSM}
		\end{center}
		\end{table}

 \begin{table}[h!!]
 \begin{center}
			\renewcommand{\arraystretch}{1.00}
			\begin{tabular}{|c|c|c|}
				\hline
				 Sector   &     Matter   & $SU(3+1)\times SU(2)_L\times SU(2)_R$  \\
				\hline\hline
				$(a,b)$ &  $ F_L $  &   $3(3+1,2,1) $  \\ 
				\hline
				$(a,c)$ &  $ F_R $  &   $3(3+1,1,2) $  \\ 
				\hline
				$(b,c)$ &  $  H $  &   $ (1,2,2) $  \\ 
				\hline
				$(a,h_1)$ &        &   $6({\bar 3}+1,1,1) $  \\ 
				\hline
				$(a,h_2')$ &   &   $6(3+1,1,1) $  \\ 
				\hline
				$(b,h_1)$ &        &   $8(1,2,1) $  \\ 
				\hline
				$(b,h_2)$ &        &   $6(1,2,1) $  \\ 
				\hline
				$(c,h_1)$ &        &   $6(1,1,2) $  \\ 
				\hline
				$(c,h_2)$ &        &   $8(1,1,2) $  \\ 
				\hline
							\end{tabular}
			\caption{Non-single chiral multiplets at the intersections of the D6 wrapping numbers of table 
            \ref{tablaSM}. They correspond to three generations
			of quarks and leptons, a Higgs multiplet and a number of extra leptons..}
			\label{espectro}
		\end{center}
		\end{table}

  This model \cite{Marchesano:2004xz} is well known in the literature,  where 
 several properties like Yukawa couplings \cite{Cremades:2004wa}
 or SUSY-breaking soft terms \cite{Font:2004cx}
 have been analysed.  If we consider now fluxes as in the first model
  in section \ref{section6},  in which all moduli are potentially fixed, chosing e.g. $h_0,q=8$ and $m,a=2$ and $N_f=8$ all RR tadpoles cancel.
  Unfortunately, as we said, fluxes must all be multiples of ,8 so this is not possible. The simplest no-scale flux choice within this class 
  of models is setting only $m,h_0\not=0$ so that,
  \beq
  W_{flux} \ = \ h_0S \  - \ m T_1T_2T_3 \ .
  \eeq
 Setting $h_0,m=8$, $N_f=8$ all tadpoles are cancelled.   There is a minimum at which the axions of $S,T_i$ vanish, and there is  one constraint
 between the saxions
 \beq
  s \ = \frac {m}{h_0} t_1t_2t_3 \ ,
  \eeq
  so that there is only one relationship among moduli.
     For the gravitino mass one has $m_{3/2}^2=(h_0m)/(32u_1u_2u_3)$.  The one-loop potentials will set the c.s. saxions $u_i$
     to their minima $\sim {\cal O}(1)$ . So this is an example of  a model  with the c.s. fields fixed at  $u_i\sim 1$, one linear combination of
     K\"ahler/ complex dilaton fixed and MSSM-like spectrum in Minkowski.  The gauge kinetic functions of the 
     MSSM sector in this model are given by
     \beq 
     2\pi f_{3+1}\ = \ 9S\ +\ U_1 \ ,\ 2\pi f_L\ =\ \frac {1}{2}U_2 \ +\ 2\pi f_R\ =\ \frac {1}{2}U_3 \ .
     \eeq
      The $SU(3)$ coupling constant may be made perturbative by having a sufficiently large vev for $s$.
     The other couplings look non-perturbative at the \textit{desert points} minima. However, those are the values
     at the string scale. The gauge groups 
     in the model are not asymptotically free, with positive $\beta$-function coefficients 
     $(b_3,b_L,b_R)=(3,9,9)$, so that, after running according to the
     renormalisation group, in the infrared all MSSM couplings become perturbative.

     Note that in all these toroidal models the gravitino mass is close to the string scale because there are very few fluxes and $W_0\simeq {\cal O}(1)$.
     In theories with many more fluxes, one should be able to obtain models with $m_{3/2}\ll M_s$.

\bibliographystyle{JHEP2015}
\bibliography{bibliography}

\end{document}